\documentclass[a4paper,11pt,preprintnumbers]{article}
\usepackage{fullpage}
\usepackage[T1]{fontenc} 
\ifx\pdfoutput\undefined
\usepackage[dvips,bookmarks]{hyperref}
\else
\usepackage{hyperref}
\fi
\hypersetup{colorlinks=false,bookmarksopen,bookmarksnumbered,citecolor=blue,
   pdfstartview=FitH}

\usepackage{latexsym}
\usepackage{amssymb,amsfonts,amsmath}
\usepackage{graphicx} 
\usepackage{indentfirst}
\usepackage{bbm}
\usepackage{amssymb}
\usepackage{verbatim}
\usepackage{amsmath, amsthm,amssymb}
\usepackage{mathrsfs}
\usepackage{hyperref}
\usepackage{amsfonts}
\usepackage{dsfont}
\usepackage{slashed, tensor}
\usepackage{booktabs}
\usepackage{graphicx}
\usepackage{mathrsfs}

\allowbreak

\newcommand{\be}{\begin{eqnarray}}
\newcommand{\ee}{\end{eqnarray}}

\def\a{{\alpha}}      
\def\b{{\beta}}
\def\g{{\gamma}}
\def\d{{\delta}}
\def\r{{\rho}}

\def\e{{\epsilon}}

\def\th{{\theta}}
\def\S{{\Sigma}}

\def\z{{\zeta}}
\def\N{{\cal N}}

\newcommand{\ri}{{\rm i}}

\def\ad{{\dot{\alpha}}}  
\def\bd{{\dot{\beta}}}

\def\ed{{\overline{\epsilon}}}
\def\thd{{\overline{\theta}}}
\def\Sd{{\overline{\Sigma}}}

\def\U{\Upsilon}

\def\eb{\boldsymbol{\eta}}

\def\D{{\rm D}}
\def\Dd{{\overline{\rm D}}}
\def\Q{{\rm Q}}
\def\Qd{{\overline{\rm Q}}}
\def\pa{\partial}

\def\Dc{{\cal D}}
\def\Dcd{{\overline{\cal D}}}
\def\Qc{{\cal Q}}
\def\Qcd{{\overline{\cal Q}}}

\def\tna{{\nabla}}
\def\btna{\overline{\nabla}}

\def\tD{\Delta}
\def\btD{\overline{\Delta}}

\def\nn{\nonumber}

\numberwithin{equation}{section}

\usepackage{xcolor}

\usepackage{cite}

\usepackage{float}

\allowdisplaybreaks

\allowbreak

\newcommand{\bea}{\begin{eqnarray}}
\newcommand{\eea}{\end{eqnarray}}

\newcommand{\1}{{\underline{1}}}
\newcommand{\2}{{\underline{2}}}

\def\U{\Upsilon}

\def\a{{\alpha}}      
\def\b{{\beta}}
\def\g{{\gamma}}
\def\d{{\delta}}
\def\r{{\rho}}

\def\e{{\epsilon}}

\def\th{{\theta}}
\def\S{{\Sigma}}

\def\z{{\zeta}}

\newcommand{\q}{{\theta}}

\def\ad{{\dot{\alpha}}}  
\def\bd{{\dot{\beta}}}

\def\ed{{\overline{\epsilon}}}
\def\thd{{\overline{\theta}}}
\def\Sd{{\overline{\Sigma}}}

\def\D{{\rm D}} 
\def\Dd{{\overline{\rm D}}}
\def\Q{{\rm Q}}
\def\Qd{{\overline{\rm Q}}}
\def\pa{\partial}

\newcommand {\cF}{{\cal F}}
\newcommand {\cG}{{\cal G}}

\newcommand {\cK}{{\cal K}}

\newcommand {\cN}{{\cal N}}
\newcommand {\cO}{{\cal O}}

\newcommand {\cU}{{\cal U}}

\def\nn{\nonumber}

\newcommand{\bsubeq}{\begin{subequations}}
\newcommand{\esubeq}{\end{subequations}}
\newcommand{\doubar}{{{\vert}\!{\vert}}}

\begin{document} 
	\begin{titlepage}
	\thispagestyle{empty}
	\begin{flushright}
	
	\end{flushright}

	\vspace{30pt}
	
	\begin{center}
	    { \Large{\bf Partial $ \cN=2$ Supersymmetry Breaking\\\vspace{0.3cm}
	     and Deformed Hypermultiplets}}
		
		\vspace{30pt}

		Fotis Farakos$^a$, Pavel Ko\v{c}\'{i}$^b$, Gabriele Tartaglino-Mazzucchelli$^{a,c}$ 
		and Rikard von Unge$^b$

		\vspace{20pt}

		{
			$^a$ {\it  KU Leuven, Institute for Theoretical Physics, \\
			Celestijnenlaan 200D, B-3001 Leuven, Belgium}

		\vspace{15pt}

			$^b${\it   Institute for Theoretical Physics, Masaryk University, \\  
			611 37 Brno, Czech Republic}

		}

		\vspace{15pt}

			$^c${\it Albert Einstein Center for Fundamental Physics,
Institute for Theoretical Physics,\\
University of Bern,
Sidlerstrasse 5, CH-3012 Bern, Switzerland}
\vspace{2mm}
~\\
Email: \texttt{
fotios.farakos@kuleuven.be,
pavelkoci@mail.muni.cz,
gtm@itp.unibe.ch,
unge@physics.muni.cz
}

		\vspace{35pt}

		{ABSTRACT}
	\end{center}
	
	\vspace{5pt}
	
We study partial supersymmetry breaking from ${\cal N}=2$ to ${\cal N}=1$ by adding non-linear terms to the ${\cal N}=2$ supersymmetry transformations. 
By exploiting the necessary existence of a deformed supersymmetry algebra for partial breaking to occur, 
we systematically use ${\cal N}=2$ projective superspace with central charges to provide a streamlined setup. 
For deformed $\cO(2)$ and $\cO(4)$ hypermultiplets, 
besides reproducing known results, 
we describe new models exhibiting partial supersymmetry breaking with and  without higher-derivative interactions.

\bigskip

\end{titlepage}

\tableofcontents

\baselineskip 6 mm

\section{Introduction}
\label{INTRO}

Despite the constraints arising from the algebra of the supersymmetry generators, 
partial supersymmetry breaking is possible \cite{Hughes:1986dn,Hughes:1986fa}. 
For ${\cal N}=2$ spontaneously broken
to ${\cal N}=1$ there are essentially two possibilities for the supersymmetry breaking sector. 
Either all the component fields of the full $\cN=2$ supermultiplets remain in the theory, or, 
some of the component fields decouple 
(by acquiring a large mass for example) 
and the remaining components form $\cN=1$ multiplets under the unbroken supersymmetry. In the latter case, 
the broken supersymmetry acts non-linearly 
on the remaining $\cN=1$ components.
Models containing complete ${\cal N}=2$ vector multiplets \cite{Antoniadis:1995vb,IZ}, 
and models with complete ${\cal N}=2$ tensor multiplets \cite{Antoniadis:2017jsk,Kuzenko:2017gsc} 
are known to exhibit  partial breaking of ${\cal N}=2$ to ${\cal N}=1$. 
Alternatively, non-linear realization techniques, including nilpotent Goldstone multiplet analysis,
 can be used to describe partial supersymmetry breaking
with only one supersymmetry manifestly preserved and linearly realized
\cite{Antoniadis:2017jsk,Bagger:1996wp,Bagger:1997pi,Rocek:1997hi,GonzalezRey:1998kh,Ambrosetti:2009za,Kuzenko:2015rfx,Kuzenko:2017gsc,Ferrara:2014oka,Dudas:2017sbi}. 
Models of this sort can be constructed from ${\cal N}=1$ chiral, vector or linear multiplets. 
The two approaches for partial supersymmetry breaking 
with complete or truncated ${\cal N}=2$ supermultiplets 
can in principle be related to one another by decoupling heavy ${\cal N}=1$ supermultiplets. 
The non-linear realization of partial supersymmetry breaking also naturally takes place 
in theories with supersymmetric extended objects (like membranes), 
which lead to supersymmetric DBI-type actions in the static gauge \cite{DP,Cecotti:1986gb}. 
Moreover, partial supersymmetry breaking is motivated from phenomenology as it can allow for a breaking of the 
extended supersymmetries in some high-energy scale while allowing a single ${\cal N}=1$
supersymmetry  in the low energy,
see e.g., \cite{Antoniadis:2012cg}. 
Along this line it is also natural, although non-trivial,
to lift partial breaking to supergravity \cite{Ferrara:1995gu,Ferrara:1995xi,Fre,Louis1,Louis2,Andrianopoli:2015wqa,Andrianopoli:2015rpa,Antoniadis:2018blk,Freedman:2012zz}.

A prerequisite for partial global supersymmetry breaking to occur is the existence of a deformed 
extended supersymmetry algebra possessing a spontaneously broken central charge symmetry (see for example \cite{Hughes:1986dn,Hughes:1986fa,Bagger:1997me,Ivanov:2000nk}). 
This property will be a guiding principle for the analysis of our work. 
More specifically, we will focus on theories possessing partial supersymmetry breaking in four dimensions
where the ${\cal N}=1$ Goldstone multiplet includes a fermion and scalar degrees of freedom.
In this case, 
under the unbroken $\cN=1$ supersymmetry, one scalar would transform as 
\be
\delta_\e \phi = \epsilon \psi + \cdots  \, , 
\label{t1}
\ee
where the fermion $\psi$ is the goldstino of the broken supersymmetry,
therefore we would have  
\be
\delta_\r \psi = f \r + \cdots   \, ,
\label{t2}
\ee
and  $f$  the constant supersymmetry breaking scale.
In this paper we will denote with the constant spinor parameter $\e$ the first, typically unbroken, supersymmetry 
while the constant spinor $\r$ will parametrize the second supersymmetry.
With the transformations \eqref{t1} and \eqref{t2}, the closure of the supersymmetry algebra would require 
$\phi$ to be the Goldstone mode of a broken central charge symmetry
generated by a scalar operator $\text{Z}$
\be
[ \delta_{\epsilon} \, , \delta_{\rho} ] \phi = \epsilon\rho\, f + \cdots = \text{Z} \phi \, . 
\ee
By analyzing different structures for spontaneously broken $\cN=2$ central charge symmetries
in supersymmetric multiplets, one could in principle classify different patterns 
for partial supersymmetry breaking.
This will be the starting point of our work.
In particular, 
by employing a projective superspace formalism,
we will describe  $4D$ $\cN=2$ supersymmetry and complete ${\cal N}=2$ matter multiplets with central charges.

Projective superspace is a formalism developed to describe theories with eight real supercharges in a manifestly 
off-shell supersymmetric way.
The main idea is based on extending the standard $\cN=2$ Minkowski superspace
${\mathbb M}^{4|8}$ 
to ${\mathbb M}^{4|8} \times {\mathbb C}P^1$ 
where the auxiliary sphere allows to efficiently organize general supersymmetric multiplets
in terms of so-called projective superfields \cite{KLR,LR1,LR2,G-RRWLvU}.\footnote{The superspace 
${\mathbb M}^{4|8} \times {\mathbb C}P^1$ was introduced for the first time  by Rosly \cite{Rosly}.
The same superspace is at the heart of the, closely related, harmonic  \cite{GIKOS,GIOS}
and projective \cite{KLR,LR1,LR2} superspace approaches.} 
These, besides being functions of the coordinates of ${\mathbb M}^{4|8}$,
also depend holomorphically on a complex parameter $\z$ 
which is an inhomogeneous coordinate of ${\mathbb C}P^1$.\footnote{More precisely projective
superfields are required in general to be holomorphic over an open domain of ${\mathbb C}P^1$.}
This formalism has been used to study manifestly supersymmetric hyper-K\"ahler sigma-models,
see \cite{Lindstrom:2008gs,Kuzenko:2010bd} for reviews,
Yang-Mills multiplets \cite{LR2,GonzalezRey:1997db,Davgadorj:2017ezp}, 
and recently also developed for an off-shell covariant description of general supergravity-matter couplings
with eight real supercharges in various dimensions 
\cite{ProjectiveSugra5D,ProjectiveSugra4D,ProjectiveSugra2D,ProjectiveSugra3D,ProjectiveSugra6D,Projective-Dan-1,Projective-Dan-2}.

By employing projective superspace, 
and focusing on partial supersymmetry breaking arising from ${\cal N}=2$ hypermultiplets, 
we will study models 
 based on real $\cO(2)$  \cite{Siegel:1978yi,Lindstrom:1983rt,N=2tensor,SSW,LR1}
  and $\cO(4)$ \cite{SSW,HST-Superactions}  hypermultiplets. 
The bosonic sector of the real $\cO(2)$ multiplet\footnote{The $\cO(2)$ multiplet is mostly known in the literature as
the $\cN=2$ linear or tensor multiplet.} 
is described by one complex scalar, 
one real scalar and one real two-form \cite{Lindstrom:1983rt}. 
These fields are contained inside two ${\cal N}=2$ superfields $\bf\Phi$ and $\bf G$ 
that lead to a chiral and a real linear superfield when reducing to ${\cal N}=1$ components. 
When the partial breaking takes place they appear with deformed supersymmetry transformations
linked to a non-vanishing central charge. 
The deformed $\cO(2)$ multiplet, 
which we denote with ${\bf H}(\z)$, possesses the following expansion in the parameter $\z$ 
\be
{\bf H}(\z) 
= \frac{{\bf \overline{\Phi}}}{\z} + {\bf G} - \z  {\bf \Phi} 
\, .
\label{O(2)-first-time}
\ee
The main idea of our paper is to parametrize possible deformations of the supersymmetry 
by the action of the 
complex central charge symmetry generator $\text{Z}$ on ${\bf H}$ as follows
\be 
\text{Z} \, {\bf H}(\z) = \frac{\a}{\z} + \b - \z  \g 
~,
\label{ZH}
\ee
where $\a$, $\b$ and $\g$  are complex constants. 
In this way, 
in the models we will consider, 
the central charge symmetry is spontaneously broken and in addition the supersymmetry can be partially broken. 
By employing a superspace with central charge, 
the ${\cal N}=2$ supersymmetry transformations will include  additional symmetry breaking terms 
arising from \eqref{ZH}. 
The complex constants $\a$, $\b$ and $\g$ are introduced to parametrize such supersymmetry breaking. 
We will show that the case where $\a=\g=0$ and $\b\ne0$ is equivalent to the 
deformation studied in 
\cite{Bagger:1996wp,Rocek:1997hi,GonzalezRey:1998kh,Antoniadis:2017jsk,Kuzenko:2017gsc}
while the other cases are new. 
In particular, we will first study these new cases in more detail in an ${\cal N}=1$ setup,
since they have not been investigated before, 
and subsequently we will introduce the projective superspace formalism. 
In a similar way we will then deform the $\cO(4)$ multiplet obtaining new models for partial supersymmetry
breaking together with reproducing the construction of \cite{GonzalezRey:1998kh}.

Once we have a new manifest ${\cal N}=2$ superspace description we will 
construct various actions which exhibit 
partial supersymmetry breaking. 
The possibility to have nontrivially interacting two-derivative models will depend on the Goldstone mode for the
 central charge symmetry.
We will see that even in cases where the two derivative theory is free due to a residual shift symmetry of the scalars, 
supersymmetric higher-derivative interactions can still be introduced, 
which are constructed directly from ${\cal N}=2$ superspace. 
Therefore, the manifest superspace description we propose here 
not only paves the way to uncover a variety of partial supersymmetry breaking patterns, 
but also for the construction of the possible interactions. 
In this work we will restrict our study to the $\cO(2)$ and the $\cO(4)$ hypermultiplets, 
but we believe that our method can also be extended and applied to other multiplets as well.

The paper is organized as follows.
In section \ref{REAL}  we review known results
for partial supersymmetry breaking based on tensor multiplets and we introduce new types of deformations
and models by using an $\cN=1$ superspace approach.
In section \ref{PRJ} we set up a projective superspace approach where deformed hypermultiplets
are linked to patterns  of central charge symmetry breaking in an $\cN=2$ superspace. 
Within this approach, here we also describe how to construct actions invariant under the deformed
$\cN=2$ supersymmetry.
Section \ref{O2} and \ref{O4} are devoted to the construction of new models for
partial supersymmetry breaking by starting from 
$\cO(2)$ and $\cO(4)$ multiplets. We will present two derivative and higher-derivative theories.
In section \ref{Discussion} we conclude by discussing our results and possible future research directions.
We accompany the paper with three appendices. 
In appendix \ref{AppendixA} we discuss properties of the deformed linear multiplet introduced in 
\cite{Kuzenko:2017oni} which naturally arise from our discussion in section \ref{REAL}.
Appendix \ref{AppendixB} includes some comments about the structure of supercurrents
 in the case of one of our models for partial supersymmetry breaking.
Appendix \ref{AppendixC} elaborates on the relation of an $\cO(4)$ model of section \ref{O4} with one
of the $\cO(2)$ models of section \ref{O2}.

\section{Deformed tensor multiplet and partial supersymmetry breaking}  
\label{REAL}

In this section we  present a new class of models 
describing partial supersymmetry breaking
based on a 
deformed 
${\cal N}=2$ tensor, or $\cO(2)$, multiplet.
The 
models we discuss here are in some sense complementary to the ones already known in the 
literature 
\cite{Bagger:1996wp,Rocek:1997hi,GonzalezRey:1998kh,Antoniadis:2017jsk,Kuzenko:2017gsc}.
There partial supersymmetry breaking is realized in terms of a Goldstone tensor multiplet
where the goldstino transforms to a real scalar and a two-form field under the unbroken supersymmetry 
transformation.  
Here we will show that in the new model 
the goldstino transforms to a complex scalar field belonging to an $\cN=1$ chiral multiplet
under the unbroken supersymmetry.  
In this section we  present our results by using an 
${\cal N}=1$ superspace formalism 
postponing the ${\cal N}=2$ superspace analysis to the second
 part of the paper. 
The purpose of this section is to present in a simple way 
the physical aspects of the new tensor multiplet models
and clarify the similarities and differences with previously known constructions. 
Building on the lessons learned in this section we will generalize and construct new models in the following sections.

An ${\cal N}=2$ tensor multiplet in 
${\cal N}=1$ superspace 
is given in terms of a chiral superfield and a real linear superfield \cite{Lindstrom:1983rt}. 
The ${\cal N} =1$ chiral multiplet is described by a superfield $\Phi$ satisfying the differential constraint
\be
\label{DefChiral}
\overline \D_{\dot \alpha} \Phi = 0 \, .
\ee
Here $\overline \D_{\dot \alpha}$ is one of the $\cN=1$ superspace covariant derivatives
$\D_A=(\D_\a,\overline \D_{\dot \alpha},\pa_{\a\ad})$, where
\be
\D_\alpha = \partial_\alpha + \frac{\ri}{2} \overline \theta^{\dot \alpha} \partial_{\alpha \dot \alpha}  
\, , 
\quad 
\Dd_\ad = \partial_{\dot \alpha} + \frac{\ri}{2} \theta^\alpha \partial_{\alpha \dot \alpha}  \, , 
\label{N=1derivatives}
\ee 
satisfying the 
only non-vanishing (anti)commutator 
\be
\{ \D_\alpha ,  \overline \D_{\dot \alpha} \} =\ri \, \partial_{\alpha \dot \alpha} 
\, .
\ee
Note that for $\cN=1$ superspace we closely follow the notations and conventions of \cite{SUPERSPACE}. 
The component fields of the chiral multiplet 
are defined as 
\be 
\label{componentsPhi}
\Phi | = A \, ,
\quad \D_{\a} \Phi | =  \chi_{\a} \, ,
\quad \D^2 \Phi | =  F \, ,
\ee
where a vertical bar next to a superfield 
indicates the projection to its $\theta=\bar\theta=0$ component, {\em i.e.}
$U(x,\theta,\bar\theta)|\equiv U(x,\theta,\bar\theta)|_{\q=\bar\q=0}$.

The ${\cal N}=1$ real linear ($\cN=1$ tensor) superfield is defined by the following constraints
\be
\overline \D^2 G = 0 = \D^2 G \, , \quad G = \overline G \, . 
\ee
Its component fields are 
\be
\label{Gcomp}
G| = \varphi \, , \quad \D_\alpha G| = \psi_\alpha \, , \quad \frac12 [\D_\alpha , \overline \D_{\dot \alpha} ] G | 
=  h_{\alpha \dot \alpha} \, , 
\ee
where $\varphi$ is a real scalar, and the field $h_{\alpha \dot \alpha}$ is the Hodge-dual of the field strength of 
a real two-form $B_{ab}$, 
$h_a=\varepsilon_{abcd}\pa^bB^{cd}$,
and as such it satisfies $\partial^{\alpha \dot \alpha} h_{\alpha \dot \alpha} = 0$. 

On both the chiral and real linear superfields
the first supersymmetry acts in the standard covariant way, 
namely\footnote{In this paper we make a conventional choice in the definition
of the $\cN=1$ supersymmetry transformations
with an opposite sign compared to the one of \cite{SUPERSPACE}.} 
\bea
\label{STSUSY}
\delta_\epsilon U 
=-\ri \epsilon^\alpha {\rm Q}_\alpha U 
 -\ri \overline \epsilon^{\dot \alpha} \overline {\rm Q}_{\dot \alpha}U 
= \epsilon^\alpha \D_\alpha U 
+ \overline \epsilon^{\dot \alpha} \overline \D_{\dot \alpha} U 
- \ri \big ( \e^\a \overline \theta^\ad + \ed^\ad \theta^\a  \big) \pa_{\a \ad}  U 
\, , 
\eea
where
\be
{\rm Q}_\alpha = \ri\partial_\alpha +\frac{1}{2} \overline \theta^{\dot \alpha} \partial_{\alpha \dot \alpha}  
\, , 
\quad 
{\overline{\rm Q}}_\ad = \ri\partial_{\dot \alpha} +\frac{1}{2} \theta^\alpha \partial_{\alpha \dot \alpha}  \, , 
\label{Q_N=1}
\ee 
are the $\cN=1$ global supersymmetry charges.
The second supersymmetry transformations of the $\cN=2$ tensor multiplet, 
which mix the $\cN=1$ chiral and real linear 
multiplets, are given by \cite{Lindstrom:1983rt}
\be 
\label{SS2} 
\delta_\rho \Phi =  - \overline  {\rho}^{\ad} \overline \D_{\ad} G \, , 
\quad 
\delta_\rho G =   \rho^{\a} \D_{\a}  \Phi 
+ \overline  {\rho}^{\ad} \overline \D_{\ad} \overline \Phi \,  . 
\ee
These supersymmetry transformations close off-shell. 

To avoid any possible confusion 
in indicating the first and the second supersymmetry, 
we will follow the convention where the $\e$-supersymmetry is always associated with the $\N=1$ superspace and 
the $\rho$-supersymmetry is the second supersymmetry transforming ${\cal N} =1$ multiplets into each other.

The most general ${\cal N}=1$ Lagrangian up to two derivatives which gives an invariant action
under the transformations 
\eqref{SS2} is  
\be 
\label{LLLL1}
{\cal L}_G = \int d^4 \theta \, {\cal H}(\Phi, \overline \Phi, G) + 
\left[\, \tilde m^2 \int d^2 \theta \, \Phi + {\rm c.c.} \right] \, , 
\ee 
where the function ${\cal H}(\Phi, \overline \Phi, G)$ satisfies the three-dimensional Laplace equation
\be
\frac{\partial^2 {\cal H}}{\partial G^2} +  \frac{\partial^2 {\cal H}}{\partial \Phi \partial \overline \Phi} = 0 \, , 
\ee 
and $\tilde m$ is a complex constant. 
On-shell the ${\cal N}=2$ tensor multiplet describes the same degrees of freedom as the ${\cal N}=2$ hypermultiplet, 
however, 
due to the presence of the two-form gauge field, the former multiplet is always massless. 
More properties of the $\cN=2$ tensor multiplet can be found in \cite{Lindstrom:1983rt}. 
We will comment on the $\cN=2$ superspace realization of this model in section \ref{O2}.
It is not difficult to show that, if $\tilde{m}^2=0$, the vacuum structure of the model \eqref{LLLL1} 
preserves the full $\cN=2$ supersymmetry.

Now we turn to describe partial supersymmetry breaking from ${\cal N}=2$ to ${\cal N}=1$. 
The way we will approach the partial breaking in this section 
is by deforming the definitions of the ${\cal N}=1$ multiplets and their transformation laws \eqref{SS2} under the 
$\rho$-supersymmetry. 
There are two simple possibilities how this can be implemented: 
\begin{enumerate}

\item We consistently deform the action of the $\rho$-supersymmetry when it acts on $G$. This has been presented 
in \cite{Antoniadis:2017jsk,Kuzenko:2017gsc},
elaborating on the nilpotent Goldstone models of \cite{Rocek:1997hi,GonzalezRey:1998kh}.  

\item We consistently deform the action of the $\rho$-supersymmetry when it acts on $\Phi$. 
This is a new deformation which we will present here in detail. 
We will see that there are two ways to achieve this, both leading to the same physics. 

\end{enumerate}
Later, in section \ref{PRJ}, we will explain the underlying mechanism behind both of these deformations
in a full $\cN=2$ superspace approach.

Before describing the new deformation we review the model presented in \cite{Antoniadis:2017jsk}. 
In that case the deformation is implemented in the $\rho$-supersymmetry by adding a term to the transformation 
of the real linear superfield $\delta_\text{def.} G = \tilde M^2 ( \theta \rho + \overline \theta \, \overline \rho )$ where 
$\tilde M^2$ is an arbitrary constant, while leaving the transformation of the chiral superfield untouched \eqref{SS2}.
The complete deformed $\rho$-supersymmetry transformations are
\be 
\label{SS2'} 
\delta_\rho \Phi =  - \overline  {\rho}^{\ad} \overline \D_{\ad} G \, , 
\quad 
\delta_\rho G =   \rho^{\a} \D_{\a}  \Phi 
+ \overline  {\rho}^{\ad} \overline \D_{\ad} \overline \Phi 
+ \tilde M^2 ( \theta^\alpha \rho_\alpha + \overline \theta^{\dot \alpha} \overline \rho_{\dot \alpha} ) 
\,  . 
\ee
The $\e$-supersymmetry is then preserved while the $\rho$-supersymmetry is broken spontaneously, 
and the goldstino is the fermion $\psi_\alpha$ defined in \eqref{Gcomp}. 
Therefore, together with the real scalar $\varphi$  and the gauge two-from $B_{ab}$,
 the goldstino forms an $\cN=1$  linear multiplet under the unbroken supersymmetry.
The generic Lagrangian with up to two derivatives which leads to an action invariant under 
\eqref{SS2'} is given by \cite{Antoniadis:2017jsk}
\be 
\label{ANT}
\begin{aligned}
{\cal L} = &\int d^4 \theta \left[  \Phi  \overline W (\overline \Phi) +  \overline \Phi W(\Phi)  
- \frac12 G^2 \left( W'(\Phi) + \overline W'(\overline \Phi)  \right)  \right]
\\
& + \left[ \int d^2 \theta \, \left (  \, \tilde m^2 \, \Phi - \tilde M^2   W(\Phi) \right) + {\rm c.c.}   \right] \, . 
\end{aligned}
\ee
Consistent propagation requires Re$W'>0$. 
As explained in \cite{Antoniadis:2017jsk}, 
$\e$-supersymmetric vacua with partial supersymmetry breaking of the $\r$-supersymmetry
exist only for both non-vanishing 
$\tilde m$ and $\tilde M$, and one has to ask that $W'' \ne 0$. 

It is worth mentioning that, in the undeformed case with $\tilde m^2=\tilde M^2=0$,
 the Lagrangian \eqref{ANT}, which is a special case of the general 
 self-interacting $\cN=2$ 
tensor multiplet 
Lagrangian \eqref{LLLL1},
was given in \cite{GHK}, where  a projective superspace derivation of the rigid $c$-map 
construction \cite{Cecotti:1988qn} was obtained.

Let us now turn to the new deformations. 
We deform the transformation of the chiral superfield under the $\rho$-supersymmetry
keeping untouched the $\rho$-supersymmetry of the linear multiplet. 
There are essentially two ways to introduce a constant deformation of the  
transformations of the chiral multiplet.
Namely 
\be
\label{ChiralDef}
\delta_\text{def.1} \Phi = - f  \, \overline {\rho}^{\ad} \overline  \theta_{\ad}  
\, , \quad 
\delta_\text{def.2} \Phi = - \tilde f \, {\rho}^{\a}  \theta_{\a} \, ,
\ee 
where $f$ and $\tilde f$ are constants that, for convenience, we choose to be real.

We start by considering the first deformation in \eqref{ChiralDef}.  
The deformed $\rho$ transformations consequently read
\be 
\begin{aligned}
\label{S2}
\delta_\rho \Phi &= - \overline {\rho}^{\ad} \overline \D_{\ad} L \, - f  \, \overline {\rho}^{\ad} \overline  \theta_{\ad}  \, , \\
\quad 
\delta_\rho L &=  \rho^{\a} \D_{\a} \Phi + \overline {\rho}^{\ad} \overline \D_{\ad} \overline \Phi  \, ,
\end{aligned}
\ee
whereas the $\epsilon$-supersymmetry transformations of the deformed $\cN=2$ tensor multiplet 
are not modified and are given by 
\be
\begin{aligned}
\label{SS111}
\delta_\epsilon \Phi &= \epsilon^\alpha \D_\alpha \Phi - \ri \big( \e^\a \overline \theta^\ad + \ed^\ad \theta^\a  \big)
 \pa_{\a \ad} \Phi \, , \\ 
\delta_\epsilon L &= 
 \epsilon^{\a} \D_{\a}  L + \overline  {\epsilon}^{\ad} \overline \D_{\ad} L 
 - \ri \big( \e^\a \overline \theta^\ad + \ed^\ad \theta^\a  \big) \pa_{\a \ad} L 
\, .
\end{aligned}
\ee
In \eqref{S2} we have deformed the action of the $\rho$-supersymmetry of the chiral superfield by an explicit 
$\overline \theta_{\dot \alpha}$ term.
To preserve the chirality constraint of $\Phi$, eq.\,\eqref{DefChiral},
 and therefore the $\cN=1$ supersymmetry transformations \eqref{SS111},
one has to deform the linear multiplet constraint as follows
\be
\label{DefModReal}
\overline \D^2 L = f = \D^2 L \, .
\ee 
Note that this constraint was recently introduced in \cite{Kuzenko:2017oni} to study 
linear multiplet models for $\cN=1$ 
supersymmetry breaking.
In our discussion this constraint naturally appears in an $\cN=2$ context.
It is straightforward to check that, thanks to \eqref{DefModReal}, we have 
\be
\overline \D_{\dot \beta} \left( \overline {\rho}^{\ad} \overline \D_{\ad} L \, 
+ f  \, \overline {\rho}^{\ad} \overline  \theta_{\ad} \right) = 0 \, ,
\ee
and then $\delta_\rho \Phi $ is chiral,
namely $\overline\D_\ad\delta_\rho \Phi =0$.

The component fields of the deformed real linear multiplet $L$ are defined as in \eqref{Gcomp}, 
where $L$ is used instead of $G$, 
and with the difference that the component $\D^2 L|$ is now a real constant instead of being zero. 
To avoid any confusion with the standard real linear multiplet 
we will define the components of the deformed real linear multiplet as 
\be
\label{Lcomp} 
L| = l \, , \quad \D_\alpha L| = \lambda_\alpha \, , \quad \D^2 L| = f
 \, , \quad 
\frac12 [\D_\alpha , \overline \D_{\dot \alpha} ] L | =  t_{\alpha \dot \alpha} \, , 
\ee 
with $t_{\a\ad}$, as $h_{\a\ad}$ in \eqref{Gcomp}, being the Hodge-dual of the field strength of a real two-form. 
We refer the reader to \cite{Kuzenko:2017oni} and appendix A of our paper for more  
properties of the deformed real linear multiplet.

We can now write down the supersymmetry transformations of the component fields. 
The component fields  of the ${\cal N}=1$ chiral superfield transform under the $\epsilon$-supersymmetry as 
\be
\label{QS1}
\delta_{\epsilon} A =  \epsilon^{\a} \chi_{\a} \, , \quad  
\delta_{\epsilon} \chi_\a =  -\epsilon_{\a} F + \ri  \overline \epsilon^{\ad} \pa_{\a \ad} A \, , \quad  
\delta_{\epsilon} F = -\ri  \overline \epsilon^{\ad} \pa_{\a \ad} \chi^\a \, , 
\ee
while  
for the components of the deformed real linear multiplet
we have  
\be
\label{QS2} 
\delta_{\epsilon} l =   \epsilon^{\a} \lambda_{\a} + \overline \epsilon^{\ad} \overline \lambda_{\ad}  , \ \    
\delta_{\epsilon} \lambda_\a  
= - \epsilon_{\a} f - \overline \epsilon^{\ad} \left ( t_{\a \ad} - \frac{\ri}{2} \pa_{\a \ad} l \right )  , \ \    
\delta_{\epsilon} t_{\a \ad } =  \frac{\ri}{2} \epsilon_{(\a}  \pa^{\b}_{\ \ad}\lambda_{\b)}
+ {\rm c.c}.  
~~~
\ee
From \eqref{QS2} we see that, assuming that in the vacuum $\langle F \rangle=0$,
 the theory contains the goldstino fermion in the deformed ${\cal N}=1$ real linear superfield, 
and it transforms under the broken $\e$-supersymmetry to the real scalar $l$ and the (Hodge-dual of the) two-form. 
In agreement with the discussion in \cite{Kuzenko:2017oni},
we see that it is the manifest $\epsilon$-supersymmetry which is broken in this setup. 
However, in components the discussion about which supersymmetry is manifest or not becomes rather academic 
since we are free to choose which of the two supersymmetries will be represented with ${\cal N}=1$ superfields.
With respect to the $\rho$-supersymmetry we find the transformations
\be
\label{QS3}
\delta_{\rho} A = -  \overline  \rho^\ad \overline \lambda_\ad \, , \quad 
\delta_{\rho} \lambda_\a =  -\rho_\a F - \ri \overline \rho^\bd \pa_{\a \bd} \overline A   \, , \quad 
\delta_{\rho} F =  -\ri  \overline \rho^{\ad} \pa_{\a \ad} \lambda^{\a}  \, , 
\ee
and
\be
\label{QS4}
\delta_{\rho} l =   \rho^\a \chi_\a + \overline \rho^\ad \overline \chi_\ad \, , \quad 
\delta_{\rho} \chi_\a =  \overline  \rho^\ad \left ( t_{\a \ad} + \frac{\ri}{2} \pa_{\a \ad} l \right ) \, , \quad  
\delta_{\rho} t_{\a \ad } = - \frac{\ri}{2} \rho_{(\a} \pa^{\b}_{\ \ad}  \chi_{\b)} 
+ {\rm c.c.} 
~,
\ee
and, again assuming $\langle F \rangle=0$, we see that the $\rho$-supersymmetry is preserved.
Notice that under the preserved supersymmetry the goldstino forms an ${\cal N}=1$ chiral multiplet. 
This shows that the new deformation is indeed different from the one discussed in \cite{Antoniadis:2017jsk}, 
where the goldstino forms a real linear multiplet under the preserved supersymmetry.

The observation that the goldstino, under the preserved supersymmetry,  
sits in the same multiplet as the complex scalar $A$ 
has a strong impact on the possible Lagrangians one can write down using this deformation of the 
$\cN=2$ tensor multiplet.  
We will elaborate more on this in section \ref{PRJ} and \ref{O2}
but the basic argument is very simple.
The goldstino is massless in global supersymmetry
and because it forms a multiplet with the complex scalar under the preserved $\rho$-supersymmetry, 
the full supersymmetric scalar multiplet which contains the goldstino has to be massless too. 
Indeed, the complex scalar $A$ has to possess a shift symmetry, which forces it to be massless, 
and ultimately  
is related to the fact that $A$  
is a goldstone mode of a spontaneously broken central charge symmetry. 
We can verify this by calculating the commutators of the supersymmetry transformations on the various fields.  
The two supersymmetries generically commute 
\be
\left [ \delta_{\epsilon}, \delta_{\rho}\right ] (\text{all fields except}\, A) = 0 \, , 
\ee
except for when they act on the complex scalar $A$, 
where it holds 
\be
\left [ \delta_{\epsilon}, \delta_{\rho}  \right ] A 
=  \overline  \epsilon_\ad \overline  \rho^\ad f \, . 
\ee
The reason this is happening is that by breaking the $\rho$-supersymmetry, 
we have also broken the central charge symmetry associated to the generator Z 
which arises from the supercharge algebra as 
\be
\{\Q_{\1\, \a} ,\Q_{\2\, \b} \} \sim C_{\alpha \beta} \text{Z} \, . 
\ee 
The goldstone mode for this  breaking associated to Z is the complex scalar $A$ 
which 
transforms as \cite{Bagger:1997me} 
\be
\text{Z} \, A = f \, . 
\ee 
This result will be the guiding principle to set up 
our analysis  
of partial supersymmetry breaking in the rest of the paper.

From the previous discussion
it is clear that in order for the deformed supersymmetry transformation to be a symmetry of a specific model 
we have to take into account the requirement that the theory has to possess a shift symmetry for the chiral superfield 
\be
\label{Sa}
\Phi \to \Phi + \text{const.} 
\ee 
The restrictions that a shift symmetry \eqref{Sa} imposes on the scalar manifold for the ${\cal N}=2$ 
single tensor multiplets were studied, e.g., in \cite{Ambrosetti:2010tu}.  
By also taking into account the invariance under the $\rho$-supersymmetry,
it turns out that for a two derivative theory the 
function ${\cal H}(\Phi, \overline \Phi, G)$ in the model \eqref{LLLL1} is constrained to be quadratic in fields.
Moreover, to ensure the $\e$-supersymmetry is not spontaneously broken on the vacuum 
by the linear superpotential, it is necessary to impose $\tilde m^2=0$.
The resulting model is described by the following simple Lagrangian 
\be 
\label{L1}
{\cal L} = \int d^4 \theta \, \Phi \overline \Phi 
+ \left( \frac14 \int d^2 \theta \, \overline{\D}_{\dot \alpha} L \,  \overline{\D}^{\dot \alpha} L  + {\rm c.c.} \right) \, . 
\ee 
In components, up to total spacetime derivatives, this reduces to 
\be
\label{LC1}
{\cal L} = \frac{1}{2}A \pa^{\a\ad} \pa_{\a \ad} \overline  A  
+ \ri \chi^{\a} \pa_{\a}^{~ \ad}   \overline  \chi_{\ad}   
+ F \overline  F 
+\frac{1}{2} t^{\a \ad} t_{\a \ad} 
+ \frac{1}{8}  l \pa^{\a \ad} \pa_{\a \ad} l 
+ \ri  \lambda^{\a} \pa_{\a}^{~ \ad} \overline  \lambda_{\ad} \, . 
\ee
This model is invariant under the supersymmetry transformations \eqref{QS1} - \eqref{QS4}, but it is a non-interacting 
theory.\footnote{In appendix B we have also explicitly calculated the supercurrents arising from \eqref{LC1}. } 
Moreover, the invariance holds for any value of the parameter $f$, which is not an observable
 in the action \eqref{LC1}. 
This is in sharp contrast compared to the model described by the action \eqref{ANT} 
which is invariant under \eqref{SS2'}. 
However, 
higher-order interactions 
that do not alter the vacuum structure of the free theory \eqref{L1}, namely $\langle F \rangle=0$, can be 
introduced.\footnote{An example of such an interaction term is given  by  \eqref{KOCIMOCI2}.}
Therefore, for this model, we should consider a Lagrangian describing an effective theory of the form 
\be 
\label{L1A}
{\cal L} = \int d^4 \theta \, \Phi \overline \Phi 
+ \left( \frac14 \int d^2 \theta \, \overline{\D}_{\dot \alpha} L \,  \overline{\D}^{\dot \alpha} L  + {\rm c.c.} \right) 
+ {\cal L}_\text{int.} \, . 
\ee 
Here ${\cal L}_\text{int.}$ is invariant under the deformed susy transformations \eqref{QS1} - \eqref{QS4} 
and may include interaction terms dependent on the parameter $f$. 
As we will see later, higher-order interactions, 
that in general might include higher-derivative terms, 
can be constructed with these properties. 
Their construction becomes straightforward by using the $\cN=2$ superspace approach
--
the framework to construct these actions is one of the main result of our paper. 
Of course the Lagrangian \eqref{L1A} will have to be treated as an effective theory 
because the higher-order terms will 
be introduced with a suppression scale $\Lambda$. 
From this point of view, and according to our analysis of the deformed supersymmetry algebra,
the action \eqref{ANT} is the only self-interacting, two-derivative QFT model for partial supersymmetry breaking based 
on deformed tensor multiplets.

Let us now elaborate on the possibility of introducing mass terms in the self-interacting Lagrangian \eqref{L1A}, 
or if mass deformations can arise from perturbative quantum corrections. 
In particular we would like to discuss the possibility of introducing mass deformations within 
any action invariant under \eqref{QS1} - \eqref{QS4}, 
assuming that the vacuum preserves $\langle F \rangle =0$. 
From the Goldstone theorem it is known that the goldstino ($\lambda$) is massless, 
but we would like to clarify what symmetries protect the other fields from getting a non-trivial mass on the vacuum. 
We will discuss the complex scalar $A$, 
the real scalar $l$, 
and the fermion $\chi$, and, 
as we will see,
there is a tight web of symmetries that keep all of the matter fields massless.  
Starting from the complex scalar $A$, 
we have already explained that under the preserved $\rho$-supersymmetry \eqref{QS3} 
it will form a chiral multiplet together with the goldstino. 
Given that the goldstino is massless on the vacuum, 
supersymmetry dictates that the complex scalar has to share this property. 
Therefore, not only can we not add mass terms for $A$, 
but in addition it has to remain massless at any order in perturbation theory, as does the goldstino. 
Now we turn to the real scalar $l$ and the fermion $\chi$. 
The two aforementioned fields form a real linear multiplet under the preserved $\rho$-supersymmetry \eqref{QS4}, 
together with a physical two-form gauge field ($t_a$ is the Hodge-dual of the field strength of the gauge two-form). 
Since the gauge two-form is always massless the same has to hold for the real scalar $l$ and the fermion $\chi$. 
This of course holds at any order in perturbation theory as well. 
The only way the gauge two-form could become massive is if it would combine via 
a ``BF'' term with a gauge abelian vector. 
However, this would require new degrees of freedom to be introduced, which are not available unless one explicitly 
includes them in the theory, 
and in addition can not arise within perturbation theory. 
We therefore conclude that, in the new model,
gauge invariance and supersymmetry protect all physical fields in \eqref{LC1} and \eqref{L1A} from receiving a mass.

We now turn to the second possibility in \eqref{ChiralDef} for deforming the $\rho$-supersymmetry on $\Phi$. 
In this case we have 
\be 
\begin{aligned}
\label{SSS2}
\delta_\rho \Phi & = - \overline {\rho}^{\ad} \overline \D_{\ad} G \, - \tilde f  \, {\rho}^{\a} \theta_{\a}  \, , 
\\
\delta_\rho G & =  \rho^{\a} \D_{\a} \Phi + \overline {\rho}^{\ad} \overline \D_{\ad} \overline \Phi  \, . 
\end{aligned}
\ee
Here the real linear multiplet is not modified
and the deformation of the $\rho$-supersymmetry of $\Phi$ preserves its chirality,
${\overline\D}_\ad\delta_\rho \Phi =0$\footnote{If instead of \eqref{SSS2} we consider the transformations
$\delta_\rho \Phi$ and  $\delta_\rho \tilde W_\ad$,
with $\tilde W_\ad:=\overline\D_\ad G$, 
the resulting variations reproduce the supersymmetry transformations
of a deformed $\cN=2$ vector multiplet in $\cN=1$ superfields. 
These were already considered  for example in one of the Goldstone models of \cite{Bagger:1997pi}.}.

It is not difficult to prove that 
the transformations \eqref{SSS2} leave the action \eqref{LLLL1} 
invariant only for 
\be
{\cal H} = \Phi \overline \Phi - G^2/2 \, , 
\ee
and one has to set $\tilde m^2 =0$ to avoid spontaneous breaking of the $\e$-supersymmetry. 
Therefore in components 
this theory has a Lagrangian of the form 
\be
\label{LC13}
{\cal L} = \frac{1}{2}A \pa^{\a\ad} \pa_{\a \ad} \overline  A  + \ri \chi^{\a} \pa_{\a}^{~ \ad}   \overline  \chi_{\ad}   
+ F \overline  F 
+\frac{1}{2} h^{\a \ad} h_{\a \ad} 
+ \frac{1}{8}  \varphi  \pa^{\a \ad}\pa_{\a \ad} \varphi 
+ \ri  \psi^{\a} \pa_{\a}^{~ \ad} \overline  \psi_{\ad} \, . 
\ee
Once we write the transformation \eqref{SSS2} in components 
we obtain for the chiral multiplet
\be
\label{QS5}
\delta_{\rho} A =  \rho^{\a} \psi_{\a} \, , \quad  
\delta_{\rho} \chi_\a =  - \rho_{\a} \tilde f - \overline \rho^{\ad} \left ( h_{\a \ad} 
- \frac{\ri}{2} \pa_{\a \ad} \varphi \right )  \, , \quad  
\delta_{\rho} F = -\ri  \overline \rho^{\ad} \pa_{\a \ad} \psi^\a \, , 
\ee
and for the component fields of the real linear multiplet  we find 
\be
\label{QS6} 
\delta_{\rho} \varphi =   \rho^{\a} \chi_{\a} + \overline \rho^{\ad} \overline \chi_{\ad}  \, , \quad  
\delta_{\rho} \psi_\a  =  -\rho_{\a} F + \ri  \overline \rho^{\ad} \pa_{\a \ad} A  \, , \quad  
\delta_{\rho} h_{\a \ad } =  \frac{\ri}{2} \rho_{(\a}\pa^{\b}_{\ \ad}\chi_{\b)} 
+ {\rm c.c.}
\ee
Clearly these transformations are the  same
as \eqref{QS1} and \eqref{QS2} with the fields of the multiplets interchanged, 
since \eqref{QS5} and \eqref{QS6} concerns the $\rho$-supersymmetry 
while \eqref{QS1} and \eqref{QS2} concerns the $\epsilon$-supersymmetry. 
The $\epsilon$-supersymmetry transformations of $\Phi$ and $G$ can be evaluated from \eqref{STSUSY}, 
and one can see that they match \eqref{QS3} and \eqref{QS4} after 
 appropriately interchanging the component fields. 
Therefore, we conclude that the two deformations in \eqref{ChiralDef} are equivalent, 
with the only difference being the interchange of the labeling of the two supersymmetries.

\section{Projective superspace, central charge and partial breaking} 
\label{PRJ}

In this section we will present in detail the properties of the ${\cal N}=2$ supersymmetry algebra with central charge 
and discuss deformed $\cO(2p)$ multiplets in projective superspace. 
We will treat the explicit examples of $\cO(2)$ and $\cO(4)$ multiplets in the next sections.

The ${\cal N}=2$ superspace with central charge is parameterized 
by the coordinates $z^M=(x^m, \th^{a\,\a}, \thd^{\ad}_{a}, z, \overline z)$. 
The $x^m$ and $\th^{a\,\a}, \thd^{\ad}_{a}$ are the standard ${\cal N}=2$ superspace coordinates, 
while $z$ and $\overline z$ are bosonic complex coordinates which we add to represent the action of the central 
charge \cite{Sohnius:1978fw}. 
Within this setup, the central charge can be written as 
\be
 \{ \D_{\1\, \a} , {\D}_{\2\, \b} \} \sim \pa_{z}~,~~~~~~
  \{ \Dd^\1_{\ad} , {\Dd}^{\2}_{\bd} \} \sim \pa_{\overline{z}} \,.
\ee
In this case a general superfield depends also on $z$ and $\overline z$. 
In contrast to the finite expansion in  $\th^{a\,\a},\thd^{\ad}_{a}$, 
the $z$ dependence has to be fixed by appropriate constraints. 
Here we will utilize such constraints to deform a  given $4D$
projective multiplet and achieve spontaneous partial breaking. 
Generic Lagrangians of the deformed multiplets would however explicitly break the ${\cal N}=2$ invariance and 
therefore one has to search for specific classes of theories that keep the ${\cal N}=2$ invariance intact and break it 
only spontaneously. 
To eliminate the $z$ dependence of the final Lagrangians we always set $z$ and $\overline z$ to zero at the end of 
the computations.

The procedure that we propose
for finding partial supersymmetry breaking models within a central charge superspace
is then the following: 
\begin{enumerate}

\item Introduce all possible central charge deformations for a given projective supermultiplet. 
This is the most important aspect of our work since in this way all possibilities for partial breaking can be 
systematically uncovered. We will focus on constant deformations.

\item Given a specified deformation, look for the most generic invariant action up to two derivatives. 
At this stage the problem is that for several deformations it is not possible to introduce interactions. 
When two-derivative interactions are possible,  we give an intuitive method in $\N=1$ superspace, 
and a general method based on projective superspace techniques is presented in subsection \ref{THETAP}.

\item Introduce higher-derivative interaction terms invariant under the central charge symmetry
and therefore invariant under the full ${\cal N}=2$ deformed supersymmetry  transformations. 
For every deformation, higher-derivative interactions can be systematically constructed
introducing a wealth of new interacting models with partial supersymmetry breaking.

\end{enumerate}

\subsection{Projective superspace with central charge}

\label{CC}

In this section we present the technical setup for our approach and also explain the general strategy proposed 
for systematically 
constructing multiplets exhibiting partial supersymmetry breaking.

The ${\cal N}=2$ superspace derivatives with a central charge 
realize the following algebra\footnote{Our conventions for the $\e^{ab}$ 
can be summarized as: $\e^{\1\2}=1\,,\,\e^{a b} = - \e^{b a}\,,\,\overline{(\e_{ab})}=\e^{ab}\,$ 
and $\e^{a b} \e_{b d} = -\delta^a_{~d}\,,\, a,b \in \1,\2$.} 
\be
\begin{aligned}
\{ \D_{a\, \a} , \Dd^{b}_{\bd}\} &= \ri \d^b_{~a}~ \pa_{\a\bd } \, ,  \\ 
 \{ \D_{a\, \a} , {\D}_{b\, \b} \} &=  \e_{ab} ~C_{\a \b}~ \pa_{z} \, , \\  
 \{ \Dd^a_{\ad} , \Dd^b_{\bd} \} &= \e^{ab} ~\overline{C}_{\ad \bd} ~\pa_{\overline{z}} \, . 
\label{n2_algebra}
\end{aligned}
\ee  
An explicit representation of the covariant derivatives is given by
\be
\D_{a\,\a} = \pa_{a\,\a} + \frac{\ri}{2} \thd^{\ad}_{a} \pa_{\a\ad} + \frac{1}{2} \e_{ba} C_{\b\a} \th^{b\,\b} \pa_{z}
\, , \quad 
\Dd^{a}_{\ad} = \pa^{a}_{\ad} + \frac{\ri}{2} \th^{a\,\a} \pa_{\a\ad} 
+ \frac{1}{2} \e^{ba} \overline{C}_{\bd\ad}  \thd^{\bd}_{b} \pa_{\overline{z}} \, , 
\label{derivatives-1}
\ee 
and a representation of the supersymmetry generators is 
\be
\Q_{a\,\a}=\ri \pa_{a\,\a} + \frac{1}{2} \thd^{\ad}_{a} \pa_{\a\ad} 
- \frac{\ri}{2} \e_{ba} C_{\b\a} \th^{b\,\b} \pa_{z} \, , \quad 
\Qd^{a}_{\,\ad}=\ri \pa^{a}_{\ad} + \frac{1}{2} \th^{a\,\a} \pa_{\a\ad}
 - \frac{\ri}{2} \e^{ba} \overline{C}_{\bd\ad}  \thd^{\bd}_{b} \pa_{\overline{z}} \, , 
\label{charges-1}
\ee 
where the supercovariant derivatives anti-commute with the supersymmetry generators as usual. 
The ${\cal N}=2$ supersymmetry transformations are defined as\be
\delta \, {\cal U} = -\ri \epsilon^{a \alpha} \, \Q_{a \alpha} \, {\cal U} 
-\ri \overline \epsilon_a^{ \dot \alpha} \, \Qd^a_{ \dot \alpha} \, {\cal U} \, , 
\label{susy-transf-1}
\ee
for some generic superfield ${\cal U}=\cU(z^M)$.

In contrast to ${\cal N}=1$ superspace, 
a simple integration over the full ${\cal N}=2$ superspace generically leads to theories with higher derivatives, 
as can be seen from dimensionality arguments. 
Therefore one has to find ways to construct invariants by integrating over only four out of the total of eight theta 
coordinates. 
One method to do this is with the use of projective superspace \cite{KLR,LR1,LR2,G-RRWLvU};
see \cite{Lindstrom:2008gs,Kuzenko:2010bd} for reviews.
In the case of projective superspace with central charge and unbroken 
$4D$ $\cN=2$ supersymmetry
the reader might look at the following papers
\cite{GonzalezRey:1997xp,Kuzenko:2006nw} and,
in the related cases of a $4D$ description with central charges of $5D$ and $6D$ multiplets,
the papers \cite{Kuzenko:2005sz,GPT-M}.\footnote{The literature on superspace techniques for 
unbroken $\cN=2$ supersymmetry with  central charge is quite ample. See for instance \cite{Dragon:1998nv}
and references therein also for an harmonic superspace description.}

It is convenient to first break the $SU(2)$ covariant notation of the ${\cal N}=2$ superspace derivatives by defining 
\be
\D_{\1 \a} \equiv \Dc_\a  \, , \quad \D_{\2 \a} \equiv \Qc_\a \, . 
\ee 
Following the notation of \cite{G-RRWLvU},
the projective covariant derivatives are  defined as\footnote{Depending on $SU(2)$ notations,
different papers use different, though equivalent, definitions for the projective superspace derivatives which,
up to complex conjugations, affect the structure of the multiplets.
The reader should use some care comparing results in the literature.
For example, in the notations used in \cite{Kuzenko:2010bd} 
an $\cO(2)$ multiplet have a chiral $\Phi$, instead of an antichiral $\overline{\Phi}$ as in \eqref{O(2)-first-time},
 $\cN=1$ superfield as the first term in its $\z$ expansion.} 
\be
\begin{aligned} 
\tna_\a(\z) =& \, \Dc_\a+\z\Qc_\a
\,, 
\\ 
\btna_\ad(\z) =& \, \Qcd_\ad- \z \, \Dcd_\ad \,,
\\
 \tD_\a(\z) =& \, -\Qc_\a + \z^{-1}  \Dc_\a
\,,
\\
\btD_\ad(\z) =& \,  \Dcd_\ad + \z^{-1} \Qcd_\ad\, , 
\end{aligned}
\ee
where $\z$ is a complex (inhomogeneous) coordinate on  the north chart of ${\mathbb C}P^1$
(see \cite{Kuzenko:2010bd} for more details about a description in terms of homogeneous, isotwistor, 
coordinates of ${\mathbb C}P^1$). 
The projective covariant derivatives realize the following vanishing anti-commutator relations 
\be
 \{ \tna_\a , \tna_\b \} = \{ \tna_\a , \btna_\bd \} =\{ \btna_\ad , \btna_\bd \} =0
 ~,~~~~~~ \{ \tD_\a , \tD_\b \} =  \{ \tD_\a , \btD_\bd \} = \{ \btD_\ad  , \btD_\bd \} = 0 \,, 
 \label{integrb-cond-proj}
\ee  
whereas the non-vanishing anti-commutators are given by  
\be
\begin{aligned}
\{ \tna_\a , \tD_\b \} = & \, - 2 C_{\a \b} ~ \pa_{z} \,,  
\\
\{ \btna_\ad , \btD_\bd \} = & \, - 2  \overline{C}_{\ad \bd} ~ \pa_{\overline{z}} \,, 
\\
\{ \tna_\a , \btD_{\ad} \} = & \, -  \{ \btna_{\ad} , \tD_\a \} =  2 \ri ~ \pa_{\a \ad} \, .  
\end{aligned}
\ee
The properties of the projective superspace derivatives under complex conjugation are 
\bsubeq
\bea
\btna_\ad(\z)= -\z \widetilde{\left( \tna_\a(\z) \right)}
&,&~~~~~~
\tna_\a(\z)= \z \widetilde{\left( \btna_\ad(\z) \right)}\,,\\
\btD_\ad(\z)= -\frac{1}{\z} \widetilde{\left( \tD_\a(\z) \right)}
&,&~~~~~~
\tD_\a(\z)= \frac{1}{\z} \widetilde{\left( \btD_\ad(\z) \right)}\,.
\eea 
\esubeq
Here the tilde conjugation is complex conjugation composed with the antipodal map on
${\mathbb C}P^1$,
which is such that $\widetilde{(\z)}\to-1/\z$, and $\widetilde{(\D_{a\,\a})}=\overline{(\D_{a\,\a})}=\Dd^a_{\ad}$.
This conjugation allows to preserve holomorphicity on the north chart of ${\mathbb C}P^1$.
For convenience, in the rest of the paper we will always indicate the tilde conjugation simply with
an overline.

By introducing the projective covariant derivatives we get two anti-commuting subalgebras which can be used to 
define invariant subspaces and supermultiplets.
A projective superfield ${\bf \Xi}={\bf \Xi}(z^M,\z)$, is an $\cN=2$ superfield which is further constrained to be
a holomorphic function of $\z$ (on an open domain of ${\mathbb C}P^1$) and to satisfy the following conditions 
\be
\tna_\a{\bf  \Xi} =0~,~~~~~~\btna_\ad{\bf  \Xi}=0\,.
\label{projconstr}
\ee 
The consistency of the previous constraints is guaranteed by the integrability conditions \eqref{integrb-cond-proj}.
It is convenient to represent the superfield ${\bf \Xi}(z^M,\z)$ by a power series in $\z$
\bea
{\bf \Xi}(z^M,\z)=\sum_{k=-\infty}^{+\infty}\z^k \,{\bf \Xi}_k(z^M)
~,
\eea
where ${\bf \Xi}_k(z^M)$ are $\cN=2$ superfields which in general might also have a dependence on the $z,\bar{z}$
central charge superspace coordinates.

It is simple to prove that, given a projective superfield ${\bf \Xi}(z^M,\z)$ and having defined its conjugate as
\bea
\overline{{\bf \Xi}}(\z):=\widetilde{({\bf \Xi}(\z))}~,~~~~~~
\overline{{\bf \Xi}}(\z)=\sum_{k=-\infty}^{+\infty}\Big(-\frac{1}{\z}\Big)^k \overline{{\bf \Xi}}_k
~,
\eea
with $\overline{{\bf \Xi}}_k(z^M)$ the complex conjugates of the $\cN=2$ superfields ${\bf \Xi}_k(z^M)$,
then $\overline{{\bf \Xi}}(\z)$ is also projective: $\tna_\a \overline{{\bf \Xi}} =\btna_\ad \overline{{\bf \Xi}}=0$.

Note that the analyticity constraints \eqref{projconstr}, rewritten in terms of the 
$(\Dc_\a,\Dcd_\ad)$ and $(\Qc_\a,\Qcd_\ad)$ derivatives, read
\bsubeq\label{projective-2}
\bea
\Qc_\a{\bf \Xi}(\z)=-\frac{1}{\z}\Dc_\a{\bf \Xi}(\z)
~~~&\Longleftrightarrow&~~~
\Qc_\a{\bf \Xi}_k=-\Dc_\a{\bf \Xi}_{k+1}
~,
\\
\Qcd_\ad{\bf \Xi}(\z)=\z\Dcd_\ad{\bf \Xi}(\z)
~~~&\Longleftrightarrow&~~~
\Qcd_\ad{\bf \Xi}_k=\Dcd_\ad{\bf \Xi}_{k-1}
~.
\eea
\esubeq
The right hand side of the previous equations can be interpreted by thinking that the dependence 
of a projective superfield upon the second superspace coordinates 
$(\theta^{\underline{2}\,\a},\bar{\theta}_{\underline{2}}^\ad)$
is completely determined in terms of the 
$(\theta^{\underline{1}\,\a},\bar{\theta}_{\underline{1}}^\ad)$ ones.
This property is the main reason why projective superspace leads to a natural description of $\cN=2$ 
supersymmetry in terms of $\cN=1$ superfields and,
as already shown in \cite{GonzalezRey:1998kh,Rocek:1997hi,Kuzenko:2017gsc} 
and as we will further see in our work,
 partial $\cN=2\to\cN=1$ supersymmetry breaking.
To elaborate more on this property
 it is worth describing the supersymmetry transformations of ${\bf \Xi}$ once reduced to
$\cN=1$ superspace. In this paper, given an $\cN=2$ superfield $\cU(z^M)$, we denote with
$\cU\doubar:=\cU|_{\theta^{\underline{2}}=\bar{\theta}_{\underline{2}}=z=\overline{z}=0}$
its reduction to the $\cN=1$ superspace parametrized by the coordinates 
$(x^m, \th^{\a}, \thd^{\ad})\equiv(x^m, \th^{\underline{1}\,\a}, \thd^{\ad}_{\underline{1}})$.
Then the supersymmetry transformations \eqref{susy-transf-1} imply\footnote{The reader should keep in mind that, 
with the definition we use for the $\cN=1$ projection $\doubar$, 
in general $ (\Qc_{\a}{\cal U})\doubar\ne\Qc_{\a}({\cal U}\doubar)$
and $(\Qcd_{\ad} {\cal U} )\doubar\ne\Qcd_{\ad} ({\cal U}\doubar)$
as well as
$(\pa_z {\cal U})\doubar\ne\pa_z ({\cal U})\doubar$
and
$(\pa_{\overline{z}} {\cal U})\doubar\ne\pa_{\overline{z}} ({\cal U})\doubar$.}
\bea
\delta_{\epsilon,\rho} \, {\cal U}\doubar 
&=&
 -\ri \big(\epsilon^{\alpha} \Q_{\alpha}+\overline \epsilon^{\dot \alpha}  \Qd_{\dot \alpha}\big)  {\cal U}\doubar
 + \r^{\alpha} \Big(\D_{\2\a}{\cal U} 
 - \th_{\a} \pa_{z}{\cal U}\Big)\doubar
+\overline \rho^{\dot \alpha} \Big( \Dd^\2_{\ad} {\cal U}
 - \thd_{\ad} \pa_{\overline{z}}{\cal U}\Big)
 \doubar 
\nn\\
&=&
\d_\e {\cal U}\doubar
+\Big(\r^{\alpha} \Qc_{\a}{\cal U} 
+\overline \rho^{\dot \alpha} \Qcd_{\ad} {\cal U} \Big)\doubar
-\Big( 
\r^{\alpha}  \th_{\a} \pa_{z}{\cal U} 
 +\overline \rho^{\dot \alpha} \thd_{\ad} \pa_{\overline{z}}{\cal U}\Big) \doubar 
 \, , 
\label{susy-transf-reduced}
\eea
where we have used \eqref{derivatives-1} and \eqref{charges-1},
 introduced the $\cN=1$ supercharges \eqref{Q_N=1}, and defined
\be
\epsilon^\a:=\epsilon^{\1\a}
~,~~~
\overline{\epsilon}^{\ad}:=\overline{\epsilon}_{\underline{1}}^\ad
~,~~~~~~
\rho^\a:=\epsilon^{\2\a}
~,~~~
\overline{\rho}^{\ad}:=\overline{\epsilon}_{\2}^\ad
~.
\ee
It is clear that the first term in \eqref{susy-transf-reduced} 
is an $\cN=1$ supersymmetry transformation of ${\cal U}\doubar$. 
Note also that the central charge 
contribution disappears from the derivatives $\D_{\1\,\a}=\Dc_\a$ and $\Dd^\1_{\ad}=\Dcd_\ad$ 
once projected to $\cN=1$ superspace,
which become precisely the standard $\cN=1$ superspace derivatives 
$(\D_\a,\Dd_\ad)$ defined in eq.\,\eqref{N=1derivatives}.
The same holds for the supercharges $\Q_{\1\,\a}$ and $\Qd^\1_\ad$
which become precisely the standard $\cN=1$ supercharges
$(\Q_\a,\Qd_\ad)$ defined in eq.\,\eqref{Q_N=1}.
On the other hand, the $(\r,\,\overline\r)$ terms 
in \eqref{susy-transf-reduced}  describe the second supersymmetry, which in general has 
central charge dependent contributions.
In the case of a projective superfield  ${\bf \Xi}$, the transformations of ${\bf \Xi}\doubar$ largely simplify. 
In fact, by using \eqref{projective-2} in \eqref{susy-transf-reduced} one obtains
\bsubeq
\bea
\delta_{\epsilon,\rho} \, {\bf \Xi}(\z)\doubar 
&=&
\d_\e {\bf \Xi}(\z)\doubar
-\frac{1}{\z}\r^{\alpha} \D_{\a}{\bf \Xi}(\z)\doubar
+\z\overline \rho^{\dot \alpha} \Dd_{\ad}{\bf \Xi}(\z)\doubar
-\r^{\alpha}  \th_{\a} \pa_{z}{\bf \Xi}(\z)\doubar
 -\overline \rho^{\dot \alpha} \thd_{\ad} \pa_{\overline{z}}{\bf \Xi}(\z)\doubar 
 \, , ~~~~~~~~~
 \\
 \delta_{\epsilon,\rho} \, \Xi_k
&=&
\d_\e \Xi_k
-\r^{\alpha} \D_{\a}\Xi_{k+1}
+\overline \rho^{\dot \alpha} \Dd_{\ad}\Xi_{k-1}
-\r^{\alpha}  \th_{\a}\text{Z}\,\Xi_k
-\overline \rho^{\dot \alpha} \thd_{\ad} \overline{\text{Z}}\,\Xi_k
 \, ,
 \label{susy-transf-reduced-2}
\eea
\esubeq
where for simplicity we have started to use the notation $\Xi_k:={\bf \Xi}_k\doubar$. We have also defined
 the action of the central charge generators $\text{Z}$ and $\overline{\text{Z}}$
on an $\cN=1$ superfield $U:={\cal U}\doubar$ reduced from an $\cN=2$ superfield ${\cal U}(z^M)$ as 
\be
\text{Z}\,U:=(\pa_z {\cal U})\doubar~,~~~~~~
\overline{\text{Z}}\,U:=(\pa_{\overline{z}} {\cal U})\doubar
~.
\ee
Differently from the general case, \eqref{susy-transf-reduced-2}
 defines a closed set of transformations among the $\cN=1$ superfield components 
 $\Xi_k$ of a projective superfield ${\bf\Xi}(\z)$.  
The supersymmetry transformation \eqref{susy-transf-reduced}
 including the central charge contributions will be the starting point when we 
describe partial supersymmetry breaking with the central charge terms producing the necessary deformations of the 
$\cN=2$ supersymmetry algebra.

In particular, we will focus on models constructed from the so-called real $\cO(2p)$ multiplets 
\cite{ProjectiveSuperspace2,G-RRWLvU}.
These are described by a projective superfield $\eb(\z)$
whose $\z$ dependence 
 can be written as 
\be
\label{O(2p)}
\eb(\z) = \sum_{k=-p}^{p} \z^k \,\eb_k
\, , 
\quad \overline 
\eb(\z) = \sum_{k=-p}^{p} \Big(- \frac{1}{\z} \Big)^k \,\overline  \eb_k 
\, , 
\ee
and the reality condition is implemented as
\be 
\eb = \overline \eb \, . 
\ee 
In this work we will carefully study two cases: $\cO(2)$ and $\cO(4)$ multiplets. 
We expect multiplets with higher $p$ to share the same supersymmetry breaking pattern as the $\cO(4)$ case since 
they contain the same physical components.
As we have already explained, 
we will make use of supermultiplets with non-vanishing central charge
that lead to partial supersymmetry breaking. 
Therefore we impose 
\be
\label{brokencharge}
\pa_{z}\eb = \sum_{k=-p}^{p}  \z^k \, \a_k
\, , \quad 
\pa_{\overline z}\eb = \sum_{k=-p}^{p} (-1)^k   \z^k\, \overline  \a_{-k} \,, 
\ee
where the $\a_i$ are complex constants. 
Equation \eqref{brokencharge} is the source of the partial supersymmetry breaking.  
Partial breaking will typically occur when only a single constant $\a_i$ is nonzero, 
otherwise if we allow for generic configurations of non-vanishing $\a_i$ supersymmetry will be 
generically (but not always) completely broken. 
A discussion about all the possibilities of switching on the $\a_i$ would be very interesting but it is
 beyond the scope of our work. 
We will therefore ask that only a single $\a_i$ at a time is non-vanishing and using the phase rotation of  
the central charge  we can choose it   to be real,\footnote{Let us assume that $\a_j = |\a_j| e^{i \b}$. 
Then if we perform a phase rotation on the central charge coordinate $z$ as $z^{\prime} = e^{-\ri\b}z$, 
\eqref{brokencharge} becomes $\pa_{z^{\prime}} \eb =  \z^j |\a_j|$. However, this procedure works if only one of the 
deformation parameters is nonzero. In the general case the $\a_i$ are complex.}
namely 
\be
\a_j \neq 0 
\, , \quad 
\a_j = \overline \a_j
\, , \quad 
\a_{i \ne j} = 0 
\, . 
\label{3.22}
\ee 
This procedure is a new proposal that unifies the description of different models of partial supersymmetry breaking 
using $\cN =2$ scalar multiplets 
and  it can be used to systematically construct 
the required ${\cal N}=2$ deformed supermultiplets.
In the next sections we will apply this philosophy to the $\cO(2)$ and $\cO(4)$ cases 
and see how all known results are indeed reproduced for the $\cO(2)$. 
When we reduce the projective superfields to ${\cal N}=1$ components, 
the superfield equations \eqref{brokencharge} 
will identify the appropriate shifts under the spontaneously broken central charge symmetry
and will uniquely help to identify the goldstone bosons. 
Using this setup we also construct $\cO(4)$ models which exhibit partial breaking and
which have not been constructed before. We will see how they relate to the $\cO(2)$ models. 
We expect that a similar
procedure can also be used for other ${\cal N}=2$ multiplets. We comment more about this in the discussion section
and leave such an analysis for future work.

\subsection{Action in projective superspace} 
\label{THETAP}

Generically, 
for projective superfields one can introduce the invariant action in the form\footnote{In all subsequent formulas, 
except otherwise stated,
a vertical bar
$\vert$ next to an $\cN=2$ superfield ${\cal U}(z^M)$ indicates that we are projecting to zero the Grassmann 
$\q^a$ and $\overline{\q}_a$ together with the central charge 
$z,\,\overline z$ coordinates: 
 ${\cal U}\vert:={\cal U}\vert_{\theta^a={\overline{\theta}}_a=z=\overline{z}=0}$.}
\be
\label{proj}
S = \frac{1}{ 32 \pi \ri}\int d^4 x \oint_C {d \z } \;  
 \z  \tD^2 \btD^2  \, {\cal K}(\eb,\z) \Big \vert \,,
\ee 
where $C$ is a contour in the $\z$ plane. 
In standard four-dimensional projective superspace the central charge vanishes
 and the action \eqref{proj} can be easily proven to be ${\cal N}=2$ supersymmetric. 
Of course in our case this action is not invariant for an arbitrary function $\cal K$ because of 
the non-vanishing central charges. 
Nevertheless, we can still use actions of the form \eqref{proj} to construct invariants. 
To see how this can be done, 
let us first study the form of the action \eqref{proj} in the presence of a central charge. 
To this end, it is useful to rewrite the action in terms of $\cN=1$ superspace derivatives. 
Using the identities
\be
\tD_\a(\z) = \frac{1}{\z} \left (  2\Dc_\a -   \tna_\a \right ) 
\, , \quad 
\btD_\ad(\z) = 2 \Dcd_\ad + \frac{1}{\z}  \btna_\ad 
\, , 
\ee
and 
\be
\{ \tna^\a , \Dc_\a \} = 2 \z ~ \pa_{z} 
\, , \quad 
\{ \btna^\ad , \Dcd_\ad \} =  2   \pa_{\overline{z}}  
\, , 
\ee 
together with the projectivity of $\cal K(\eb,\z)$, 
namely $\nabla_\a{\cal K}=\overline{\nabla}_\ad{\cal K}=0$,
the action \eqref{proj} takes the form
\bea
\label{actionN1}
{1 \over 2 \pi \ri} \int d^4 x\oint_C {d \z } 
\frac{1}{\z} \Big \{ \D^2 \Dd^2  {\cal K}(\eb,\z) 
+  \frac{1}{2 \z}   \pa_{\overline{z}} \,  \D^2 {\cal K}(\eb,\z)   
- \frac{1}{2} \z  \pa_{z} \Dd^2 {\cal K}(\eb,\z)  
- \frac{1}{4}  \pa_{z}\pa_{\overline{z}} {\cal K}(\eb,\z)    \Big \}  \Big \vert \, . 
~~~
\ee
The action arising from \eqref{actionN1} generically 
explicitly breaks $\r$-supersymmetry
 and, due to the last three terms,
 also the $\epsilon$-supersymmetry. 
To restore the ${\cal N}=2$ invariance of \eqref{actionN1} 
we modify the ansatz (\ref{proj}) by introducing explicit theta terms. 
The new ansatz is given in terms of a function ${\cal G}$ and reads
\be
\begin{aligned}
\label{ansatz}
{\cal G}(\eb,\z) =&  \, {\cal K}(\eb,\z) + \theta^2_p~ {\cal R}(\eb,\z) 
+ \frac{1}{\zeta^2} \overline \theta^2_p ~\overline{{\cal R}\left(\eb, \z \right)}
\\
& + \frac{1}{\zeta^2} \theta^2_p \overline \theta^2_p ~\left [ 4  \pa_{z}\pa_{\overline{z}} {\cal K}(\eb,\z) 
- 2 \z  \pa_{\overline{z}}{\cal R}(\eb,\z)  + \frac{2}{\z} \pa_{z}\overline{{\cal R}\left(\eb, \z \right)}  \right ] \, , 
\end{aligned}
\ee 
where we have defined 
\be
\label{newtheta}
\theta^\a_p = \frac{\z}{2} \left ( \theta^{\1 \a}-\frac{1}{\z}\theta^{\2 \a} \right) \, , \quad 
\overline \theta^\ad_p = \frac{1}{2} \left ( \overline \theta_{\1}^\ad+\z \overline \theta_{\2}^\ad \right) \, . 
\ee
The reason for giving the $\theta_p$ and $\overline \theta_p$ their specific form \eqref{newtheta} is that they are 
both annihilated by $\tna$ and $\btna$. 
Replacing $\cK$ with $\cG$ in the action \eqref{actionN1}
we get 
\be
\begin{aligned}
\label{newaction}
S = &\, {1 \over 32 \pi \ri}\int d^4 x \; \oint_C {d \z } \; 
 \z  \tD^2 \btD^2  \, {\cal G}(\eb,\z)\Big \vert 
 \\
  = &\, {1 \over 2 \pi \ri}\int d^4 x \; \oint_C {d \z } \; 
 \frac{1}{\z} \Big \{ \D^2 \Dd^2 {\cal K}(\eb,\z) 
+ \D^2  {\cal A} 
+\Dd^2 \overline{\cal A}   \Big \}\Big \vert \, , 
\end{aligned}
\ee
where
\be
\label{calA}
{\cal A} = \frac{1}{2} \z^{-1}  \pa_{\overline{z}} \, {\cal K}(\eb,\z) + \frac{1}{4\z^2} \overline{{\cal R}\left(\eb, \z \right)}\,.
\ee

By performing general $\cN=2$ supersymmetry transformations on the action and requiring them to vanish
\be
\begin{aligned}
\label{karlovyvary}
\d_{\epsilon,\rho} S 
 =&\,{1 \over 2 \pi \ri}\int d^4 x \; \oint_C \frac{d \z}{\z} \; 
   \Bigg \{ ~   \overline{\e}^\ad   \D^2 \Dd_{\ad} {\cal A} 
  +\overline{\rho}^{\ad}    \D^2 \Dd_{\ad} \big( \z {\cal A} - \pa_{\overline z} {\cal K}(\eb,\z) \big)\\
  & \hspace{21ex}  + \e^\a \Dd^2 \D_\a ~ \overline{\cal A}   
  + \rho^\a  \Dd^2 \D_\a ~\left (-\frac{1}{\z}  \overline{\cal A}  - \pa_{z} {\cal K}(\eb,\z) \right ) \\
 &\hspace{21ex}  - \rho^{\a} \D_\a \pa_z {\cal A} - \overline \rho^{\ad} \Dd_{\ad} \pa_{\overline z} \overline{\cal A}  
  \Bigg \}\Big \vert   
  = 0\,,
\end{aligned}
\ee
we arrive at a series of conditions on ${\cal A}$ which once satisfied lead to ${\cal N}=2$ supersymmetric theories. 
The conditions read 
\be
\label{condA}
\begin{aligned}
\e^\a:& \oint_C \frac{d \z}{\z} \; \Dd^2 \D_\a  \overline{\cal A}  = \text{total derivative}
~~~ \Longrightarrow~~~
  \oint_C \frac{d \z}{\z} \;   \overline{\cal A} = \text{anti-chiral} \, , 
\\
\rho^\a:&\oint_C \frac{d \z}{\z} \; \left[ \Dd^2 \D_\a \left (\frac{1}{\z}  \overline{\cal A}  + \pa_{z} {\cal K}(\eb,\z) \right ) 
+ \D_\a \pa_z {\cal A} \right]  = \text{total derivative} \, . 
\end{aligned}
\ee
Note that in principle $\overline {\cal A}$ could  also be a complex linear superfield but since in that case 
$\overline {\cal A}$  would completely drop out from \eqref{newaction}  we do not consider this option. 
When solving the equations \eqref{condA} it is useful to rewrite them as
\be
\label{condARikard}
\begin{aligned}
\int d^4 x& \;  \oint_C \frac{d \z}{\z} \; \D^2 \Dd_\ad \left ( \frac{1}{\z} \pa_{\overline z} {\cal K} 
+ \frac{1}{2 \z^2} \overline{\cal R} \right )  = 0 \, , 
\\
\int d^4 x& \;  \oint_C \frac{d \z}{\z} \; \D^2 \Dd_\ad \left (  \pa_{\overline z}{\cal K}
 - \frac{1}{2 \z} \overline{\cal R} \right )  = 0 \, ,\\
\int d^4 x& \;  \oint_C \frac{d \z}{\z} \;  \D_\a \left ( \frac{1}{\z} \pa_{\overline z} \pa_{z} {\cal K} 
+ \frac{1}{2 \z^2} \pa_{z}\overline{\cal R} \right )  = 0 \, .
\end{aligned}
\ee
One should not expect to find a solution for the previous conditions for all deformations.
However, by using this method,
it will be possible to construct the known two derivative theories, together with new models.
In contrast, we will also show that for any deformation 
it is straightforward to construct higher-derivative terms based on a projective superspace approach.

The procedure we described in this subsection 
is essentially equivalent to writing down ${\cal N}=1$ terms and then explicitly checking the 
invariance of the $\rho$-supersymmetry, 
while adding appropriate compensating terms. 
However, 
having now an $\cN=2$ superspace description of this procedure significantly adds to its understanding. 
Moreover, in some cases this procedure can help us tell right away 
which starting Lagrangians are bound to fail purely from projective superspace arguments. Note that the shift 
symmetry that is essential in the $\N=1$ superspace construction of the deformed theory is not used in 
the projective superspace program. 
Rather it is implied by the supersymmetry conditions \eqref{condA}. 
The methods are therefore complementary to each other which might be useful in more complicated situations.

To exemplify the projective superspace procedure we can work on a free theory for the real $\cO(2p)$ 
multiplet \eqref{O(2p)}, 
which is deformed as shown in \eqref{brokencharge}. 
We can for example start with 
\be
\label{freeaction}
S = {(-1)^p \over 64 \pi \ri}\int d^4 x \; \oint_C {d \z } \; 
 \z  \tD^2 \btD^2  \, \eb^2
 \Big \vert    \,,
\ee 
and using \eqref{karlovyvary} we find the variation of the action under ${\cal N}=2$ 
supersymmetry
\be
\label{dfree} 
\begin{aligned}
\d_{\epsilon,\rho} S  
 = ~\frac{(-1)^p \a_j }{2}\int  d^4 x \; 
   \Big \{ &(-1)^j   \,  \overline{\e}^\ad   \D^2 \Dd_{\ad} \eta_{j+1} 
  + (-1)^{j+1} \overline \a_j \, \overline{\rho}^{\ad}    \D^2 \Dd_{\ad} \eta_{j} 
  \\
 & +  (-1)^j  \, \e^\a \Dd^2 \D_\a \eta_{-j-1}    
  + (-1)^{j+1}  \, \rho^\a  \Dd^2 \D_\a \eta_{-j} \Big \} 
\Big\vert\!\Big\vert     \,.
\end{aligned}
\ee
In general, 
the variation \eqref{dfree} is not vanishing but one could possibly switch on $\cal R$ terms in the form 
of \eqref{ansatz},  to make it vanish. 
Let us now see under which circumstances it is possible to find such $\cal R$ functions. 
The real $\cO(2p)$ multiplet with $p>1$
contains a set of unconstrained auxilliary $\cN=1$ superfields $\eta_i$ for $i \in \langle -(p-2),(p-2)\rangle$ as well
as the physical chiral and complex linear superfields (we will see this later in more details for the  $\cO(4)$
cases). 
If $\eta_{j}$ in \eqref{dfree} is an unconstrained superfield
we cannot find $\cal R$ in order to render the action $\N =2$ invariant. 
In this case we would have to somehow change the ansatz in \eqref{freeaction}.
If on the other hand the $\eta_{j}$ appearing in \eqref{dfree} are chiral superfields 
the action is invariant under the full deformed supersymmetry
without any need for further manipulations. 
The most interesting case is when $\eta_i$ is a (anti-)complex linear superfield,
$\Dd^2\eta_i=0$ ($\D^2\eta_i=0$).
Since the transformation does not vanish we need to choose an $\cal R$  of the form
\be
\label{geta2}
\begin{aligned}
{\cal R}_{\eb^2}\left(\eb,\z\right) &= (-1)^p \z^{-1} \pa_{z} \, \eb^2 \, ,  &\text{for}& \ i=-p+1 \, ,\\
{\cal R}_{\eb^2}\left(\eb,\z\right) &= (-1)^{(p+1)} \z^{-1} \pa_{z} \, \eb^2 \, ,  &\text{for}& \ i=p-1\, .
\end{aligned} 
\ee 
Therefore, the free action is given by \eqref{newaction} with  
\be
{\cal K}= (-1)^p\frac{\eb^2}{2} \, , \quad {\cal R}={\cal R}_{\eb^2} \, .
\ee 
For the simpler $\cO(2)$ multiplet the quadratic action \eqref{freeaction} is always invariant.

In contrast to Lagrangians for kinetic terms and interaction terms with at most two derivatives, 
higher-derivative interactions can be constructed generically. 
For a real function ${\cal F}$, 
we can have 
\bsubeq
\label{SAA}
\bea
S_{\rm int.} &=& {1 \over 32 \pi \ri}\int d^4 x  \oint_C {d \z } \; 
 \z  \tD^2 \btD^2  \, \left [ \frac{1}{\z^2} \tna^2 \btna^2 {\cal F} \Big{(}   \eb_i, (\tD) \btD \eb_k, (\tna) \btna \eb_l,\z \Big{)}
  \right ] \Big \vert   
\\
&=&{1 \over 4 \pi \ri}\int d^4 x  d^4\theta \oint_C\frac{{d \z }}{\z} \; 
\big( \Qc^2 \Qcd^2+\Qcd^2\Qc^2\big)   \,   {\cal F} \Big{(}   \eb_i, (\tD) \btD \eb_k, (\tna) \btna \eb_l,\z \Big{)}
\Big\vert\!\Big\vert
      , \quad i \neq \pm j \, .~~~~~~~~~~~~
      \label{SAA-b}
\eea
\esubeq
The Lagrangian density is by construction projective due to the $\tna^2 \btna^2$ operator. 
Note  that, the previous action could be written as an integral over all the eight Grassmann variables of
$\cN=2$ superspace.
The condition $i \neq \pm j$ is chosen so that the component field on which the central charge generator acts 
nontrivially does not appear without a derivative in order for the resulting Lagrangian to have the required shift 
symmetry, see \eqref{3.22}.
Alternatively,
 it is sufficient to have the Lagrangian $\cF$ to be an $\cN=2$ superfield annihilated by the central charges.
Explicitly we have 
\be
\partial_z \eb_i = \partial_{\overline z} \eb_i = 0 , \, \quad \text{for} \ i \neq \pm j \, , 
\ee 
and 
\be
\partial_z \btD \eb_k = \partial_z \tD \eb_k  = 0 = \partial_z \btna \eb_k 
= \partial_z \tna \eb_k , \, \quad \text{for any} \ k \, . 
\ee 
Note also that, if $\overline{{\cal F}}={\cal F}$ then the action \eqref{SAA} is real 
and the $\Qc^2\Qcd^2$ and $\Qcd^2\Qc^2$ terms lead to the same contribution up to total derivatives.
It is a well-known fact that higher-derivative interactions may lead to ghost excitations. We have not investigated this 
further in our paper. Instead we will illustrate the general construction given in \eqref{SAA} on some simple examples 
for the $\cO(2)$ and $\cO(4)$ multiplets. A full analysis of the vacuum structure induced by these terms and a 
possible classification of ghost-free higher-derivative interactions of models with partial supersymmetry breaking is 
beyond the scope of this paper. 
The presented examples,
and possible ghost-free $\cN=2$ models,
 can be viewed as $\N=2$ extensions of the models studied in 
 \cite{Khoury:2010gb,Farakos:2013zsa,Farakos:2014iwa,Farakos:2015vba,Farakos:2016zam,Cecotti:1986jy,Gates:1995fx,Gates:1996cq,Koehn:2015scs,Nitta:2014pwa,Ciupke:2015msa,Aoki:2014pna,Koehn:2012ar,Fujimori:2016udq}.

\section{The $\cO(2)$ multiplet} 
\label{O2}

In this section we revisit the multiplet studied in section \ref{REAL}, namely the $\cO(2)$ multiplet, 
and see how it fits into the general setup presented in section \ref{PRJ}. 
Our method can reproduce all the deformations we discussed in section \ref{REAL}.

The real $\cO(2)$ multiplet is constructed by setting 
\be
{\bf H}(x, \z , z, \overline{z})= \frac{{\bf \overline{\Phi}}}{\z} + {\bf G} - \z  {\bf \Phi} \, , 
\ee
where ${\bf \Phi}$ and ${\bf G}$ are ${\cal N}=2$ superfields with ${\bf G} = \overline{\bf G}$. 
Using the projectivity condition 
\be
 \tna_\a {\bf H} = 0 = \btna_{\ad} {\bf H} \, , 
\label{constraints}
\ee
we can derive a series of constraints for the ${\cal N}=2$ superfields, 
in terms of the ${\cal N}=2$ superspace derivatives. 
We find chirality constraints on ${\bf \Phi}$, namely 
\be
\begin{aligned}
\label{projconst02A} 
\Dc_\a\overline{\bf{\Phi}} = 0 \, , \quad 
\Dcd_\ad{\bf{\Phi}} =0 \, , \quad 
\Qc_\a{\bf{\Phi}} = 0 \, , \quad 
\Qcd_\ad \overline{\bf{\Phi}} = 0 \, , 
\end{aligned}
\ee
which mean that when we reduce to ${\cal N}=1$ superspace $\Phi$ will always become a chiral superfield. 
We also find constraints that link the superspace derivatives of the ${\cal N}=2$ superfields to each other 
\be
\label{projconst02B} 
\begin{aligned}
\Qc_\a{\bf G} =\Dc_\a{\bf \Phi}  \, , \quad 
\Qcd_\ad{\bf G} =\Dcd_\ad\overline{\bf \Phi}  \, , \quad 
\Qc_\a\overline{\bf \Phi}=-\Dc_\a{\bf G} \, , \quad 
\Qcd_\ad\bf \Phi=-\Dcd_\ad{\bf G} \, , 
\end{aligned}
\ee
which will help when we reduce to ${\cal N}=1$ superspace. 
However, 
because of the central charge the constraint that normally would show that the ${\cal N}=2$ 
superfield ${\bf G}$ becomes a real linear superfield now is deformed by explicit central charge terms.
The details depend on how the central charge acts on ${\bf H}$.

Following the discussion in section \ref{PRJ}, 
the $\cO(2)$ multiplet can be deformed in three different ways, 
depending on how the central charge acts on the projective superfield ${\bf H}$. 
We set 
\be
\label{charge02}
\pa_{z} {\bf H} = \frac{\a}{\z} - \b - \z  \g \, , \quad 
\pa_{\overline{z}} {\bf H} =  \frac{\g}{\z} -  \b - \z \, \a \, , 
\ee
where $\a,\b,\g$ are real constants. 
Once we combine \eqref{charge02} with the constraints \eqref{projconst02A} and \eqref{projconst02B}, 
we find the deformed constraints of the ${\cal N}=2$ superfields, 
namely 
\be
\label{deform02}
\begin{aligned}
\Dc^2 {\bf G} & = \a \, , 
\\
\Dc^2{\bf \Phi} + \Qc^2 \overline{\bf \Phi} & = \beta  \, , 
\\
\Qc^2 {\bf G} & =\gamma \, , 
\end{aligned}
\ee
and their complex conjugates given by 
\be
\Dcd^2 {\bf G}= \a \, , \quad 
\Dcd^2 \overline{\bf \Phi} + \Qcd^2 {\bf \Phi} =   \beta \, , \quad 
\Qcd^2 {\bf G}= \gamma \, . 
\ee 
The simplest way to understand the meaning of the conditions \eqref{deform02} is to turn
 to component fields as we will do now.

Having at hand the complete set of defining constraints for the ${\cal N}=2$ superfields allows us to 
straightforwardly
define their  component fields.
A consistent definition of a part of the independent components of ${\bf H}$ 
in terms of the components of ${\bf G}$ is given by 
\be
\label{MLcompGG}
{\bf G} | = \varphi \, ,
\quad \Dc_\a {\bf G} | =  \psi_{\a} \, , 
\quad  \frac{1}{2} \, [\Dc_\a, \Dcd_\ad] {\bf G} | =  h_{\a \ad} \,  ,
\quad \Dc^2 {\bf G}|=\a \, ,
 \quad  \Qcd^2 {\bf G} | =  \g \, , 
\ee
while the other independent component fields of ${\bf H}$ can be defined in terms of components of ${\bf \Phi}$, 
namely 
\be
\label{MLcompFF}
 {\bf \Phi} | = A \, , 
\quad \Dc_\a {\bf \Phi} | =  \chi_{\a} \, ,
\quad \Dc^2 {\bf \Phi} | =   F \, , 
\quad \Qc^2 \overline{\bf \Phi} | =  \b - F \, . 
\ee 
In \eqref{MLcompGG} the field $\varphi$ is a real scalar and $h_a$ is the Hodge-dual of a real two-form, 
while in \eqref{MLcompFF} the fields $A$ and $F$ are complex scalars. 
For the component fields in \eqref{MLcompGG} the ${\cal N}=2$ supersymmetry transformations are given by 
\be
\label{SUSYG}
\begin{aligned}
\delta_{\e,\rho} \varphi &=   \epsilon^{\a} \psi_{\a}  + \rho^\a \chi_\a + {\rm c.c.} \, , 
\\ 
\delta_{\e,\rho} \psi_\a &= 
-\epsilon_\a \a - \overline \epsilon^{\ad} \left ( h_{\a \ad} - \frac{\ri}{2} \pa_{\a \ad} \varphi \right ) 
+ \rho_\a \left ( \beta - F \right ) - \ri \overline{\rho}^{\ad} \pa_{\a \ad} \overline{A} \, , 
\\
\delta_{\e,\rho} h_{\a \ad }&= 
 \frac{\ri}{2}  \left (
\epsilon_{(\a} \pa^{\b}_{\ \ad} \psi_{\b)}
+ \overline \rho_{(\ad} \pa_{\a}^{\ \bd} \overline \chi_{\bd)}
\right )  + {\rm c.c.} \, , 
\end{aligned}
\ee
while for the ones in \eqref{MLcompFF} the supersymmetry transformations are given by 
\be
\label{SUSYF} 
\begin{aligned}
\delta_{\e,\rho} A &= \epsilon^{\a} \chi_{\a} - \overline  \rho^\ad \overline \psi_\ad \, , 
\\
\delta_{\e,\rho} \chi_\a &=  
-  \e_{\a} F + \ri  \overline \epsilon^{\ad} \pa_{\a \ad} A  - \rho_\a \g 
+ \overline  \rho^\ad \left ( h_{\a \ad} + \frac{\ri}{2} \pa_{\a \ad} \varphi \right ) \, , 
\\
\delta_{\e,\rho} F &= - \ri  \overline \epsilon^{\ad} \pa_{\a \ad} \chi^\a - \ri   \overline \rho^{\ad} \pa_{\a \ad} \psi^{\a} \, . 
\end{aligned}
\ee
To study the properties of \eqref{SUSYG} and \eqref{SUSYF} we will refer to the variations 
$\epsilon_\alpha$ as the $\epsilon$-supersymmetry 
and to the variations $\rho_\alpha$ as the $\rho$-supersymmetry. 
Finally, in the notations of section \ref{REAL},
the condition \eqref{charge02} also gives the shifts of the scalars under the central charge generator 
\be
\label{shifts}
\text{Z} A = \alpha \, , \quad \text{Z} \varphi = - \beta  \, , \quad \text{Z} \overline A = \gamma  \, , 
\ee
which translate to shift symmetries in the action.

By a simple inspection of \eqref{SUSYG} and \eqref{SUSYF} we can 
analyze different supersymmetry breaking patterns. 
Let us stress however that it is not the algebra alone that defines the supersymmetry breaking pattern, 
rather it is the algebra and the vacuum of the theory which are needed. 
Assuming that only one constant from the $\alpha$, $\beta$ and $\gamma$ is non-vanishing in each case, 
the three possibilities are: 
\begin{enumerate} 

\item Setting $\a \neq 0$ breaks the $\epsilon$-supersymmetry, 
and the goldstino is given by $\psi_\alpha$. 
The goldstino forms a supermultiplet under the unbroken supersymmetry with the complex scalar $A$, 
therefore $A$ has to possess a shift symmetry, 
in agreement with \eqref{shifts}.

\item Setting $\b \neq 0$ and assuming $\langle F \rangle =0$, 
leads to the breaking of the $\rho$-supersymmetry with the goldstino $\chi_\alpha$. 
The goldstino forms a supermultiplet under the $\epsilon$-supersymmetry with $\varphi$ and the two-form. 
In agreement with \eqref{shifts} the real scalar $\varphi$ has a shift symmetry. 
Note that, if $\langle F\rangle=\b\ne0$ then supersymmetry is partially broken with the $\r$-supersymmetry preserved 
and the $\e$-supersymmetry broken \cite{Antoniadis:2017jsk}.

\item Setting $\g \neq 0$ breaks the $\rho$-supersymmetry, 
and the goldstino forms a supermultiplet under the $\epsilon$-supersymmetry with the complex scalar $A$, 
which possess a shift symmetry in agreement with \eqref{shifts}.

\end{enumerate}
The cases 1$.$ and 3$.$ above have been studied in detail in section \ref{REAL}, 
and we have also explained why they describe the same physics. 
The case 2$.$ has been studied in detail in \cite{Antoniadis:2017jsk} and we only rapidly
 reviewed it section \ref{REAL}. 
Therefore we see how the general method presented here reproduces these results.

For completeness, 
it is useful to write down how the $\rho$-supersymmetry 
acts on the ${\cal N}=1$ superfields. 
These ${\cal N}=1$ superfields are defined from the ${\cal N}=2$ superfields as 
\be
{\bf \Phi}\doubar
= \Phi \, , \quad {\bf G}\doubar
= G
 \, , 
\ee
and, once we use \eqref{susy-transf-reduced-2} for an $\cO(2)$ multiplet,
 the $\rho$-supersymmetry transformations then take the form 
\be
\begin{aligned}
\delta_{\rho} \Phi &= 
- \overline \rho^{\ad} \Dd_{\ad} G  
- \g \, \rho^\alpha \theta_\alpha 
-  \a \, \overline \rho^\ad \overline \theta_\ad \, , \\
\delta_{\rho} G  &=
  \rho^{\a} \D_{\a} \Phi  
+ \overline \rho^{\ad} \Dd_{\ad} \overline \Phi  
+ \b \left ( \rho^\a \theta_\a 
+  \overline \rho^\ad \overline \theta_\ad \right )\, .
\end{aligned}
\ee
We see that these formulas match with the deformations that were conjectured in section \ref{REAL}, 
namely \eqref{SS2'} and \eqref{ChiralDef}. 
Moreover,  when $\alpha$ is non-vanishing, equation \eqref{MLcompGG} naturally gives rise to the deformed real 
linear superfield \eqref{DefModReal}.

We now turn to the possible actions these deformations allow us to construct. 
We will not study the cases $\alpha \ne 0$ and $\gamma \ne 0$ independently, as they describe the same physics
up to exchanging the $\e$- and $\r$-supersymmetries, 
therefore we will only study the $\gamma \ne 0$ and $\beta \ne 0$ cases.

We first study the case $\beta \ne 0$ ($\alpha=0=\gamma$). 
The two-derivative $\N =1$ action of the theory is 
\be 
\label{ANT2}
\begin{aligned}
S= &\int d^4x d^4 \theta \left[  \Phi  \overline W (\overline \Phi) +  \overline \Phi W(\Phi)  
- \frac12 G^2 \left( W'(\Phi) + \overline W'(\overline \Phi)  \right)  \right]
\\
& + \left[ \int d^4xd^2 \theta \, \left(  \, \tilde m^2 \, \Phi -  \beta \, W(\Phi) \right) + {\rm c.c.}   \right] \, , 
\end{aligned}
\ee
which matches \eqref{ANT} for $\tilde M^2 = \beta$. 
Since the goldstino is described by $ \D_\a G | $ the action \eqref{ANT2} 
is invariant with  respect to $G \to G + \text{const.}$ 
A simple way to construct the action \eqref{ANT2}, 
is to start with the most general two-derivative action, 
namely \eqref{LLLL1}. 
The ansatz is restricted by imposing the aforementioned shift symmetry on $G$. 
To make the result invariant under the deformed supersymmetry one has to introduce $\cN =1$ compensating terms,
in this case $(-  \beta \, \int d^4x d^2 \theta \, W(\Phi) + {\rm c.c.})$,
as first appeared in \cite{Antoniadis:2017jsk}.

As we explained in the previous section, 
methods for constructing Lagrangians within ${\cal N}=2$ projective superspace can also be used. 
We will apply these methods to re-derive the two-derivative terms and then construct higher-derivative interactions.

For the possible functions ${\cal K}$ which can be used in \eqref{ansatz} we
consider  two simple options. 
One choice is to set 
\be
\begin{aligned}
\label{ansHg}
{\cal K}({\bf H},\z) 
= & 
-\left [\frac{\overline k({\bf H})}{\z^n} + (-\z)^n k \left( {\bf H} \right )\right ] 
\\
= & 
-\left [\frac{\overline k\left ( {\bf G} + \frac{{\bf \overline{\Phi}}}{\z} - \z  {\bf \Phi} \right )}{\z^n} 
+ (-\z)^n k \left({\bf G} + \frac{{\bf \overline{\Phi}}}{\z}  - \z  {\bf \Phi}  \right )\right ] 
\, ,
\end{aligned}
\ee 
with the function $k$ analytic in its argument.
To illustrate the form of the resulting $\N =1$ Lagrangian derived from \eqref{ansHg} 
we perform the contour integral over $\zeta$ 
which gives 
\be
\label{miroslav}
\oint_C \frac{d \z}{\z}  {\cal K}({\bf H},\z) \doubar
\ \sim \ 
\sum_{j=0}^{\infty}  \overline k^{(2j+n)}(G)\, \Phi^{n+j} \overline \Phi^j + 
{\rm c.c.}
\, , 
\ee 
where $k^{(n)}(X) = \pa^{n}k(X)/\pa X^{n}$. 
Therefore this choice would lead to an ${\cal N}=1$ superspace Lagrangian starting with a term 
$\int d^4 \theta \left ( \overline k^{(n)}(G) \Phi^n +  k^{(n)}(G) \overline \Phi^n \right )$, 
and then there would follow an infinite sum of terms 
with more derivatives on $k(G)$ of the form shown in \eqref{miroslav}. 
However, a careful analysis of \eqref{condA} leads to the conclusion that the quadratic Lagrangian density provides 
the only $\N=2$ invariant theory in \eqref{ansHg} both for $\b \neq 0$ and $\g \neq 0$.

An alternative possibility is to have 
\be
\begin{aligned}
\label{ansHphi}
{\cal K}({\bf H},\z) 
= &  
-\left [\frac{\overline k(\z{\bf H})}{\z^n} 
+ (-\z)^n k \left( -\frac{{\bf H}}{\z} \right )\right ] 
\\
= & 
-\left [ \frac{\overline k\left ( {\bf \overline{\Phi}} + \z {\bf G} - \z^2  {\bf \Phi} \right )}{\z^n} 
+ (-\z)^n k \left( {\bf \Phi} - \frac{{\bf G}}{\z} - \frac{{\bf \overline{\Phi}}}{\z^2} \right )\right ] 
\, .
\end{aligned}
\ee 
Performing the contour integral gives 
\be
\label{TT22}
 \oint_C \frac{d \z}{\z}  {\cal K}({\bf H},\z)\doubar
 \ \sim \  
 \sum_{a,b:\,a+2b=n}  \overline k^{(a+b)}(\overline \Phi) ~ G^a  \Phi^b + {\rm c.c.}  
\ee 
In this case we would have an $\cN=1$ superspace Lagrangian in terms of the function 
$\overline{k}(\overline \Phi)$ and its derivatives, 
together with appropriate powers of $G$ and $\Phi$, 
as shown in \eqref{TT22}.  It can be shown that  other choices of 
${\cal K}({\bf H},\z) = -\left(\frac{\overline k(\z^m{\bf H})}{\z^n} + {\rm c.c.}\right ) $
for $m\ne -1,0,1$ can be treated as special cases of the previous ansatz.

Turning back to the deformations \eqref{charge02} for $\beta \ne 0$, 
with the central charge nontrivially acting on ${\bf G}$, 
the obvious choice for ${\cal K}$ is \eqref{ansHphi}. 
In the undeformed case, with $n=2$, this was the starting models of \cite{GHK}
and the partial supersymmetry breaking analysis of \cite{Kuzenko:2017gsc}.
To find the correct power of $\z$ we use the crucial observation that 
\be
\label{chirality}
\oint_C \frac{d \z}{\z} {\cal A} =  \text{chiral}
\,.
\ee
Let us plug \eqref{ansHphi} into \eqref{calA} with $\overline {\cal R} =0$,
which gives
\be
\label{power}
\oint_C \frac{d \z}{\z} {\cal A}|_{\overline {\cal R} =0} 
=
\oint_C \frac{d \z}{\z} 
 \frac{\overline\b}{2}\left [ \frac{ \overline k'\left({\bf \overline{\Phi}} + \z {\bf G} - \z^2  {\bf \Phi} \right )}{\z^n} 
 + (-1)^n\z^{n-2} k' \left( {\bf \Phi} - \frac{{\bf G}}{\z} - \frac{{\bf \overline{\Phi}}}{\z^2} \right )\right ]
~.
\ee
Only if $n=2$ the second term is chiral. On the other hand the first term generates a non-chiral 
part in \eqref{power}. However, this contribution can be canceled by the appropriate choice for $\cal R$. 
Then the functions 
\be
\begin{aligned}
\label{antonfgnotfull}
{\cal K}({\bf H},\z)&=-\frac{\overline k(\z{\bf H})}{\z^2} - \z^2 k \left( -\frac{{\bf H}}{\z} \right )\,,
\\
{\cal R} \left ( {\bf H},\z \right )&= \frac{2}{\z}  \pa_{z} \left [ \frac{ \overline k(\z{\bf H})}{\z^2}
 -\z^2 k \left( -\frac{{\bf H}}{\z}\right )   \right ] \, ,
\end{aligned}
\ee 
once inserted into \eqref{newaction} give a deformed $\N =2$ invariant action. The second term in ${\cal R}$ 
makes ${\cal A}$ chiral and the first term is essential for preserving the $\rho$ supersymmetry.
It is easy to see that we can also add the linear superpotential to $\cal R$ which is always 
invariant for any deformation of the $\N=2$ supersymmetry transformations. 
Finally,  we have
\be
\begin{aligned}
\label{antonfgO2}
{\cal K}({\bf H},\z)&=-\frac{\overline k(\z{\bf H})}{\z^2} - \z^2 k \left( -\frac{{\bf H}}{\z} \right )\,,
\\
{\cal R} \left ( {\bf H},\z \right )&= 
\frac{2}{\z}  \pa_{z} \left [ \frac{\overline k(\z{\bf H})}{\z^2} -\z^2 k \left( -\frac{{\bf H}}{\z}\right )   \right ] 
+4 \tilde{m}^2 \frac{{\bf H}}{\z} \, ,
\end{aligned}
\ee 
which reproduces the results of \cite{Antoniadis:2017jsk,Kuzenko:2017gsc}.
The $\N =1$ action of the theory is given by \eqref{ANT2} with $W(\Phi)=k'(\Phi)$.

The superspace methods become more important when we want to introduce higher-derivative interactions.
Following the general discussion in the previous section we will only give one simple example. 
Using the ansatz \eqref{SAA}, 
we can consider the action
\be
S_{\rm int.}={1 \over 32 \pi \ri} \frac{1}{\Lambda^4} \int d^4x  \oint_C {d \z } \; 
 \z  \tD^2 \btD^2  \, \left [
  \frac{1}{\z^2} \tna^2 \btna^2 \left ( {\bf \Phi}^2 \overline {\bf \Phi}^2  \right ) \right ] 
  \Big \vert   \, , 
\ee 
for some cut-off scale $\Lambda$, 
which gives rise to a variety of ${\cal N}=1$ interactions 
when, by using \eqref{SAA-b}, we expand in ${\cal N}=1$ superspace, 
namely 
\be
\label{inter1}
\begin{aligned}
S_{\rm int.} = \frac{1}{2\Lambda^4}  \int d^4x d^4 \theta \; 
   \Big{[} ~ 
   &  \pa^{\a\ad} \Phi \pa_{\a\ad} \Phi \overline \Phi ^2
   +4\ri \pa^{\a \ad} \Phi \Dd_\ad G \D_\a G \overline \Phi
   + 2 \b \left ( \Dd^\ad G \Dd_\ad G \overline \Phi
   +  \D^\a G \D_\a G  \Phi \right )\\
   -& 2 \D^\a G \D_\a G \Dd^2 \overline \Phi  \Phi
   - 2 \Dd^\ad G \Dd_\ad G \D^2 \Phi \overline \Phi
   + 4 \b^2 \Phi \overline \Phi
   - 4 \b  \Phi \overline \Phi \Dd^2 \overline \Phi\\
   - &4  \b  \Phi \overline \Phi \D^2  \Phi
   + 4 \Phi \overline \Phi \D^2 \Phi \Dd^2 \overline \Phi
   + \D^\a G \D_\a G \Dd^\ad G \Dd_\ad G
   +2\Phi \Box \Phi \overline \Phi ^2
   \\
     +&4\ri \Phi \overline \Phi  \pa^{\a \ad} \Dd_\ad G \D_\a G 
\Big{]} 
              +{\rm c.c.} 
\end{aligned}         
\ee 
Notice that the ${\cal N}=1$ action \eqref{inter1}
is indeed invariant under the shift $G \to G + \text{const.}$


Before turning back to the discussion of  deformations with $\b\ne 0$, it is worth comparing 
 our construction with the $\cN=2$ superspace analysis in \cite{Antoniadis:2017jsk} and \cite{Kuzenko:2017gsc}.
In \cite{Antoniadis:2017jsk} the model \eqref{ANT} and \eqref{ANT2}
was shown to arise as a particular action for a so-called ``chiral-antichiral''
(or twisted-chiral in the nomenclature used in \cite{Kuzenko:2017gsc})  $\cN=2$ superfield 
$\cal Z$. This is such that 
\be
\overline{D}_\ad^\1 {\cal Z}=0~,~~~~~~
D_{\2\,\a} {\cal Z}=0~,
\label{CAC}
\ee
where here we denoted with 
$D_{a\,\a}$ and $\overline{D}_\ad^a $ the $\cN=2$ superspace spinor derivatives without central charges
(which then coincide with $\D_{a\,\a}$ and $\overline{\D}_\ad^a $
in \eqref{derivatives-1} once we set $\pa_z=\pa_{\overline{z}}\equiv0$).
As such, $\cal Z$ contains in general $16+16$ component fields.
The superspace integral
\be
\int d^4x \,d^2\theta^\1 \,d^2\overline{\theta}_\2 \,F({\cal Z})
\label{actionCAC}
\eea
proves to be manifestly $\cN=2$ supersymmetric for any holomorphic function $F({\cal Z})$.
If ${\cal Z}$ is further constrained to be a \emph{short}  $8+8$ multiplet thanks to the extra constraints
$\overline{D}_\ad^\2 {\cal Z}\doubar=\Dd_\ad G$, 
with $G=\overline{G}$ an $\cN=1$ real linear superfield ($\Dd^2 G=\D^2 G=0$), and
$(\overline{D}^\2)^2 {{\cal Z}}\doubar=-\Dd^2\overline{\Phi}$, with $\Phi:={\cal Z}\doubar$,
then one obtains an  equivalent 
superfield description of the $\cN=2$ $\cO(2)$ multiplet together with the action \eqref{ANT2} for $\tilde{M}^2=0$
and $W(\Phi)=F'(\Phi)$.
The $\tilde{M}^2$ term, together with the deformation of the supersymmetry transformation \eqref{SS2'},
was achieved in  \cite{Antoniadis:2017jsk}
by giving a vev to $(\bar{D}^\2)^2 {{\cal Z}}\propto\tilde{M}^2$
amounting to the redefinition 
${{\cal Z}}\to{{\cal Z}}+\tilde{M}^2\overline{\theta}_\2^2$ 
which preserves the constraint \eqref{CAC}.\footnote{Another possible deformation 
preserving  \eqref{CAC} is given by ${{\cal Z}}\to{{\cal Z}}+\tilde{A}^2({\theta}^\1)^2$ 
but, up to an exchange of the first and the second supersymmetries, it proves to be equivalent 
 to the $\tilde{M}^2\overline{\theta}_\2^2$ deformation \cite{Antoniadis:2017jsk}.
Our $\b$ deformation is clearly equivalent to this case too once properly choosing what is the manifest 
$\e$-supersymmetry and the $\r$-supersymmetry.}
In \cite{Kuzenko:2017gsc}  it was shown that the action \eqref{actionCAC}
 arises as a particular case of the $\cN=2$ projective action without central charges.
 It was then shown that the model \eqref{ANT} derives from
 a projective Lagrangian given by
 ${\cal K}({\bf H},\z)=- \Big(\frac{\overline F(\z{\bf H})}{\z^2} + \z^2 F \left( -\frac{{\bf H}}{\z} \right )\Big)$
 where ${\bf H}$ possesses a nontrivial $\overline{\theta}_p^2$ vev
along the line of the analysis of \cite{GonzalezRey:1998kh,Rocek:1997hi}.
Note that the two descriptions given in \cite{Antoniadis:2017jsk}  and \cite{Kuzenko:2017gsc}
are both equivalent to our $\b\ne 0$ and $\a=\g=0$ deformations.
On the other hand, it appears  that the $\a\ne0$ or $\g\ne0$ deformations of the supersymmetry 
transformations of an $\cO(2)$ multiplet cannot be generated by a simple
spurionic $\theta$-dependent shift 
without either breaking the constraint \eqref{CAC}
or the projectivity of ${\bf H}$.
For this reason, the studies in \cite{Antoniadis:2017jsk,Kuzenko:2017gsc}
 missed these possible deformations which we achieved by directly analyzing
the role of the central charge.
It is on the other hand possible that extending the $\cO(2)$ multiplets to a relaxed-hypermultiplet
\cite{Howe:1982tm} $\a\ne0$ or $\g\ne0$ deformations might be achieved with proper spurionic terms.


Let us now come back to our approach and consider the  $\g \ne 0$ ($\alpha=0=\beta$) deformation.
In this case we know that the model has to be invariant under the shift symmetry $\Phi\to\Phi+{\rm const.}$
Models with undeformed $\cN=2$ tensor multiplets possessing such shift symmetry were considered in 
a different context in \cite{Ambrosetti:2010tu}
where it was proven that the function ${\cal H}(G,\Phi,\overline{\Phi})$ in \eqref{LLLL1}
 is constrained to be either quadratic or cubic in its $\cN=1$ superfields.
However, invariance under the second $\g$-deformed supersymmetry 
 only allows for two-derivative actions of the form 
\be 
S = \int d^4x d^4 \theta \Big{[}  \Phi \overline \Phi - \frac12 G^2 \Big{]} 
+ \Big{[} \tilde m^2 \int d^4x d^2 \theta \, \Phi + {\rm c.c.} \Big{]} \, . 
\ee 
If we also impose the vacuum to preserve the manifest $\e$-supersymmetry, 
as already discussed in section \ref{REAL}, it is necessary to impose $\tilde{m}^2=0$.
$\cN=2$ superspace methods again become very useful in finding nontrivial higher-derivative interactions. 
For the setup we have here we need to introduce interaction terms in the form of \eqref{SAA}. 
A simple example is to have the ${\cal F}$ function to depend only on ${\bf G}$, 
since ${\bf G}$ is annihilated by the central charge. 
As an illustration of possible interactions we can consider 
\be
S_{\rm int.} =  {1 \over 32 \pi \ri}\frac{1}{\Lambda^4} \int d^4x \oint_C {d \z } \; 
 \z  \tD^2 \btD^2  \, \left [
 \frac{1}{\z^2} \tna^2 \btna^2 {\bf G}^4 \right ]
 \Big \vert    \, , 
\ee 
for some cut-off scale $\Lambda$. 
The corresponding $\N =1$ action
that we can compute by using \eqref{SAA-b}, which gives rise to a variety of interactions, 
reads 
\be
\begin{aligned}
S_{\rm int.} = \frac{1}{2\Lambda^4}  \int d^4x d^4 \theta \; 
   \Big{[}  &\, 
    12 \g^2 G^2
    +12 \g G \left ( \Dd^\ad \overline \Phi \Dd_\ad \overline \Phi
   +  \D^\a \Phi \D_\a \Phi \right )
   \\
   & + 6 \D^\a \Phi \D_\a \Phi \Dd^\ad \overline \Phi \Dd_\ad \overline \Phi 
   + 24 G \D^\a \Phi \Dd^\ad \D_\a G \Dd_{\ad} \overline \Phi 
   \\ 
   & + 12\ri  G^2  \pa_{\a}^{~ \ad} \D^{\a} \Phi \Dd_{\ad} \overline \Phi + 6 G^2 \Dd^\ad \D_\a G \Dd_\ad \D^\a G 
\Big{]} 
        +{\rm c.c.}
\end{aligned}         
\ee 
It is easy to see that there is a shift symmetry $\Phi \to \Phi + \text{const.}$ as it should be, 
since here it is the chiral superfield $\Phi$ which contains the goldstino.

Another example is given by 
\be
\label{KOCIMOCI}
{\cal F} = \frac{1}{256 \Lambda^{12}} \left [ \left ( \tna + \z \tD   \right)  {\bf \Phi} \right ]^2 
\left [ \left ( \btD - \z^{-1} \btna   \right)  \overline {\bf \Phi} \right ]^2
\left [ \left ( \tna + \z \tD   \right)  {\bf G} \right ]^2 \left [ \left ( \btD - \z^{-1} \btna   \right)  {\bf G} \right ]^2 \, ,
\ee
which can be rewritten as
\be
\label{KOCIMOCI2} 
{\cal L}_\text{int.} = \frac{1}{\Lambda^{12}} \int d^8 \theta \left (\Dc {\bf \Phi} \right )^2  
 \left (\Dcd  {\bf \overline{\Phi}} \right )^2  \left (\Dc {\bf G} \right )^2  \left (\Dcd {\bf G} \right )^2 \, ,
\ee
where $\Lambda$ is again a cut-off scale.
The bosonic sector of \eqref{KOCIMOCI2} can simply be inferred to have the following structure
\be
{\cal L}_\text{int.}^{bosons} = \frac{1}{\Lambda^{12}} F^4 \overline{F}^4
+\sum_{n=0}^{3}\sum_{m=0}^{3}F^m\overline{F}^n \cO_{m,n}(\pa A,\pa \overline{A},\pa \phi,h_{a},\g) \, .
\ee
The terms in the sum are quite involved but the main property we want to stress is that it is a functional 
at least linear in derivatives of the scalar fields and of $h_a$. 
This implies that $F$ can in principle be integrated out algebraically becoming a functional of 
$(\pa A,\pa \overline{A},\pa \phi,h_{a})$
and the deformation parameter $\g$.
It is also simple to show that the model described by
the free theory \eqref{L1} together with the higher-order Lagrangian \eqref{KOCIMOCI2}
possesses a branch of solutions for $F$ that
preserve the Lorentz invariant vacuum structure, $\langle F \rangle = 0$, while introducing 
non-trivial self-interacting higher-derivative terms.

\section{The $\cO(4)$ multiplet}
\label{O4}

In this section we study partial supersymmetry breaking by using a real $\cO(4)$ multiplet, 
which comprises an ${\cal N}=1$ complex linear, an ${\cal N}=1$ chiral, 
and a real unconstrained auxiliary $\cN=1$ superfield. 
In component form this multiplet contains two complex scalars and two Weyl fermions as physical fields. 
Therefore, 
in contrast to the models of the previous sections \ref{REAL} and \ref{O2}, 
the models we present here contain only complex scalars in the bosonic sector.

Before we deform the $\cO(4)$ multiplet and induce the partial supersymmetry breaking 
we would like to rapidly present the undeformed ${\cal N}=2$ supersymmetric theory 
in terms of the constituent ${\cal N}=1$ superfields. 
These are the chiral superfield
\be
\label{thegood} 
\overline \D_{\dot \alpha} \Phi = 0 \, , 
\ee
the complex linear superfield
\be
\label{thebad}
\overline \D^2 \Sigma = 0 \, , 
\ee
and the real unconstrained superfield
\be
\label{theugly}
X = \overline X \, . 
\ee
Under ${\cal N}=1$ they transform in the standard way as shown in formula \eqref{STSUSY}. 
The $\rho$-supersymmetry transformations read 
\be
\label{SUSYO4}
\begin{aligned}
\delta_{\rho} \Sigma  &= \rho^\alpha \D_\alpha \Phi 
- \overline  \rho^{\dot \alpha} \overline \D_{\dot \alpha} X \, , 
\\
\delta_{\rho} \Phi &=  - \overline \rho^{\dot \alpha} \overline \D_{\dot \alpha} \Sigma \, , 
\\
\delta_{\rho}X  &= \rho^\alpha \D_\alpha \Sigma
 + \overline \rho^{\dot \alpha} \overline \D_{\dot \alpha} \overline \Sigma \, , 
\end{aligned}
\ee
and they close off-shell. 
An example of a 
possible ${\cal N}=2$ supersymmetric model which will be important for later discussion takes 
the form\footnote{Notice that, in contrast to the ${\cal O}(2)$ multiplet, here a linear superpotential term
 $\tilde m^2 \int d^2 \theta \, \Phi$ is not allowed because it is not invariant under \eqref{SUSYO4}.} 
\be
\label{FULL}
\begin{aligned}
S =  \int d^4x d^4 \theta \Big{[} &\, f(\Phi) \overline \Phi + \overline f(\overline \Phi) \Phi 
+ \left( f'(\Phi) + \overline f'(\overline \Phi) \right) \Big( \frac12 X^2 - \Sigma \overline \Sigma \Big)
\\
& + \frac12 X \left( f''(\Phi) \Sigma^2 + \overline f''(\overline \Phi) \overline \Sigma^2 \right) 
+ \frac{1}{4!} \Sigma^4 f'''(\Phi) + \frac{1}{4!} \overline \Sigma^4 \overline f'''(\overline \Phi) 
\\
& + 
\frac12 W'(\Phi) \Sigma^2 + \frac12 \overline W'(\overline \Phi) \overline \Sigma^2 
+ X \left( W(\Phi) + \overline W(\overline \Phi) \right) 
\Big{]} \, , 
\end{aligned}
\ee
where $W(\Phi)$ and $f(\Phi)$ are holomorphic functions of the chiral superfield $\Phi$. 
One can easily check that \eqref{FULL} is indeed invariant under \eqref{SUSYO4}. 
From \eqref{FULL} we see that the superfield $X$ has an algebraic equation of motion which allows
 us to integrate it out
and the resultant theory will contain only a chiral and a complex linear superfield
possessing ${\cal N}=2$ supersymmetry on-shell.

\subsection{The deformed $\cO(4)$ multiplet} 

\label{DO4}

To find consistent deformations of the supersymmetry transformations and the appropriate 
modifications of the multiplets, 
we will turn to projective superspace and follow the general method developed in section \ref{PRJ}.

In projective superspace the $\cO(4)$ multiplet has the form 
\be
{\bf P}(\z) 
= \frac{{\bf \overline \Phi}}{\zeta^2} + \frac{{\bf \overline \Sigma}}{\zeta} 
+ {\bf X} - \zeta {\bf \Sigma} + \zeta^2 {\bf \Phi} \, , 
\ee 
where ${\bf \Phi}$, ${\bf X}$ and ${\bf \Sigma}$ are ${\cal N}=2$ superfields. 
The projectivity and the reality conditions on ${\bf P}$ read 
\be
\tna_\alpha {\bf P}(\z) = 0 \ , \ \btna_{\dot \alpha} {\bf P}(\z) = 0 \ , \ {\bf P}(\z) = \overline {\bf P}(\z) \, . 
\ee 
Once we write these conditions in terms of the ${\cal N}=2$ superspace derivatives 
and the constituent ${\cal N}=2$ superfields 
we find a series of constraints. 
We find as usual the chirality conditions 
\be 
\Dcd_\ad {\bf  \Phi} = 0 \, , \quad 
\Qc_\a {\bf \Phi} = 0  \, , 
\ee
a reality condition 
\be
{\bf X} = {\bf \overline X}  \, , 
\ee
which makes ${\bf  X}$ a real but otherwise unconstrained superfield, 
and a series of equations linking the various ${\cal N}=2$ superfields 
through their superspace derivatives 
\be
\begin{aligned}
\Dc_\a {\bf \overline \Sigma} + \Qc_\a {\bf \overline \Phi} &= 0  \, , 
\\
\Dc_\a {\bf X} + \Qc_\a {\bf \overline \Sigma} &= 0  \, , 
\\
 \Dc_\a {\bf \Sigma} - \Qc_\a {\bf X} &= 0  \, , 
\\
\Dc_\a {\bf \Phi} - \Qc_\a {\bf \Sigma} &= 0  \, . 
\end{aligned}
\ee 
Partial supersymmetry breaking is switched on, 
together with the deformations, 
by simply imposing the central charge to act as follows 
\be
\label{ZP}
\partial_z {\bf P} = \frac{\a}{\zeta^2}   - \frac{\b}{\zeta}    + \zeta \g + \zeta^2 \mu\, , 
\ee 
where $\alpha$, 
$\beta$, $\gamma$ and $\mu$ 
are real constants.\footnote{
We could have also assumed a deformation of the form $\partial_z {\bf P} = \text{const}$. 
But, as we have seen in subsection \ref{THETAP}, such deformation does not have an $\N=2$ invariant free theory.
Therefore we do not consider this possibility further in this article.} 
Using the supersymmetry algebra including the central charge together with 
the deformation \eqref{ZP} and the previously derived constraints, 
we find that ${\bf \Sigma}$ has to satisfy the following deformed linearity conditions 
\be
\Dcd^2 {\bf  \S} = \a \, , \quad \Qc^2 {\bf  \S} = \mu \, , 
\ee
and that ${\bf X}$ and ${\bf \Phi}$ are further related to each other through the following equations 
\be
\begin{aligned}
\Dc^2 {\bf \Phi}&=\g + \Qc^2{\bf X} \, , 
\\
\Qc^2 {\bf \overline \Phi}&= \b + \Dc^2{\bf X} \, . 
\end{aligned}
\ee

Now we want to see exactly how the partial supersymmetry breaking arises from the 
supersymmetry transformations of the constituent multiplets. 
From the ${\cal N}=2$ multiplets we can reduce to the ${\cal N}=1$ multiplets 
\be
\Phi = {\bf \Phi}\doubar
\, , \quad 
\Sigma = {\bf \Sigma}\doubar
\, , \quad 
X = {\bf X}\doubar
\, , 
\ee
which satisfy the following ${\cal N}=1$ constraints 
\be
\label{D2O4} 
\Dd_{\dot \alpha} \Phi = 0 
\, , \quad 
\Dd^2 \Sigma = \alpha
\, , \quad 
X = \overline X \, . 
\ee
We see that from our procedure a deformed complex linear superfield \cite{Kuzenko:2011ti} 
has naturally appeared.\footnote{Note that the deformed complex linear constraint in \eqref{D2O4}
is a natural limit of the constraint $\Dd^2\S=T$ first introduced in \cite{Deo:1985ix},
see also \cite{TartaglinoMazzucchelli:2004vt}, 
and that is ubiquitous when one considers off-shell $\cN = 2$ sigma-models with central charge, 
see \cite{Kuzenko:2006nw}.
Here $T=T(\Phi^I)$ is typically a holomorphic function of dynamical chiral superfields $\Phi^I$.
If $\langle T\rangle={\rm const.}\ne 0$, $\cN=1$ supersymmetry is typically broken 
on the vacuum as for \eqref{D2O4}.}
The $\epsilon$-supersymmetry transformations are as usual calculated 
for each component field of the ${\cal N}=1$ superfields 
from the formula \eqref{STSUSY},
while the $\rho$-supersymmetry acts on the $\cN=1$ superfields as follows 
\be
\label{S2O4}
\begin{aligned}
\delta_{\rho} \Phi &=- \overline \rho^{\dot \alpha} \Dd_{\dot \alpha} \Sigma  - \mu \, \rho^\alpha \theta_\alpha 
-  \a \, \overline \rho^{\dot \alpha} \overline \theta_{\dot \alpha} \, , 
\\
\delta_{\rho} \Sigma  &= \rho^\alpha \D_\a \Phi
- \overline  \rho^{\dot \alpha} \Dd_{\dot \alpha} X 
 + \g \, \rho^\alpha \theta_\alpha 
+ \b \, \overline  \rho^{\dot \alpha} \overline \theta_{\dot \alpha} \, , 
\\
\delta_{\rho} X  &= \rho^\alpha \D_\a \Sigma + \overline \rho^{\dot \alpha} \Dd_{\dot \alpha} \overline \Sigma 
\, . 
\end{aligned}
\ee 
The central charge operator, 
which we interpret as a generator of a shift symmetry on the ${\cal N}=1$ Lagrangians, 
acts on the ${\cal N}=1$ superfields  as 
\be
\label{SZO4}
\text{Z} \Phi = \mu 
\, , \quad 
\text{Z} \overline \Phi = \alpha 
\, , \quad 
\text{Z} X = 0 
\, , \quad 
\text{Z} \Sigma = - \gamma 
\, , \quad 
\text{Z} \overline \Sigma = - \beta 
\, . 
\ee 
We remind the reader that we will assume that only one of the constant parameters $\alpha$, 
$\beta$, $\gamma$ or $\mu$ is switched on at a time. 
To find the supersymmetry breaking patterns, 
one could further reduce the full ${\cal N}=2$ supersymmetry transformations to component fields. 
However, using the understanding we have about partial breaking and the implications on the central charge, 
we can readily deduce the supersymmetry breaking patterns.

By  inspection of \eqref{D2O4}, \eqref{STSUSY}, \eqref{S2O4} and \eqref{SZO4}, 
we see that the possible partial supersymmetry breaking patterns 
are the following: 
\begin{enumerate} 

\item Setting $\a \neq 0$ breaks the $\epsilon$-supersymmetry and the goldstino
 is described by the component field $\D_\a {\overline \Sigma} |$. 
The goldstino forms a multiplet under the preserved supersymmetry with the complex scalar ${\Phi}|$. 
Therefore in this setup the superfield $\Phi$ has to possess the shift symmetry.

\item Setting $\b \neq 0$ and assuming $\langle \D^2X| \rangle =0$ breaks the $\rho$-supersymmetry 
with the goldstino described by $\D_\a \overline \S|$. 
The goldstino forms a complex linear supermultiplet under
 the preserved supersymmetry with the complex scalar $\S|$, 
and therefore $\Sigma$ has to possess a shift symmetry.

\item Setting $\g \neq 0$ and assuming that $\langle {\D}^2 \Phi |\rangle = \gamma$ breaks the 
$\epsilon$-supersymmetry with the goldstino 
described by $\D_\a \Phi|$. 
The goldstino forms a multiplet with the complex scalar $\S|$, 
and therefore $\S$ has to possess a shift symmetry.

\item Setting $\g \neq 0$ and assuming that $\langle \D^2 \Phi |\rangle =0$ breaks the 
$\rho$-supersymmetry with the goldstino 
described by ${\D}_\a \S|$, which is an auxiliary fermion in the two-derivative undeformed theory. 
The goldstino forms a multiplet with the complex scalar $\S|$, 
and therefore $\S$ has to possess a shift symmetry. 

\item Setting $\mu \neq 0$ breaks the $\rho$-supersymmetry, with the goldstino described by $\D_\a \Phi |$. 
The goldstino forms a chiral supermultiplet under the preserved supersymmetry with the complex scalar $\Phi|$. 
The ${\cal N}=1$ superfield $\Phi$ has to possess a shift symmetry. 

\end{enumerate} 
We have to stress that in the above classification we are assuming that the vacuum 
does indeed preserve one of the two supersymmetries. 
This is not trivial and it is model-dependent, 
as there might be auxiliary fields that get a vev in the vacuum and break supersymmetry completely or change 
the supersymmetry breaking pattern provided by the deformations introduced by the central charges.

In the next two subsections we will discuss in more detail the cases 2$.$ and 5$.$ 
We do not need to discuss the case 1$.$ and 3$.$ in full detail since they describe the same physics 
as the cases 5$.$ and 2$.$ respectively 
and they simply correspond to exchanging the $\epsilon$- with the $\rho$-supersymmetry. 
We will not discuss the case 4$.$ in detail since it requires the fermion 
$\D_\alpha \Sigma|$ to become propagating, 
which is typically an auxiliary fermion in the undeformed two-derivative theory.

Before we discuss each deformation of ${\bf P}$ in detail, 
it is instructive to look at possible candidates for the function ${\cal K}$ in \eqref{newaction}. 
Given an analytic function $k(x)$, we can consider the following three simple possibilities: 
\begin{enumerate}

\item[i.] An expansion of ${\cal K}$ around ${\bf X}$, namely 
\be
\begin{aligned}
{\cal K}({\bf P},\z)=&\,\frac{\overline k({\bf P})}{\z^n} + (-\z)^n k \left( {\bf P} \right )
\\
=&\,\frac{\overline k\left ( {\bf X} + \frac{{\bf \overline \Phi}}{\zeta^2} + \frac{{\bf \overline \Sigma}}{\zeta} 
 - \zeta {\bf \Sigma} 
+ \zeta^2 {\bf \Phi}  \right )}{\z^n} 
+ (-\z)^n k \left({\bf X} + \frac{{\bf \overline \Phi}}{\zeta^2} + \frac{{\bf \overline \Sigma}}{\zeta}  
- \zeta {\bf \Sigma} + \zeta^2 {\bf \Phi} \right ) 
\, . 
\end{aligned}
\ee

\item[ii.] An expansion of ${\cal K}$ around ${\bf \S},\overline {\bf \S}$, 
namely 
\be
\begin{aligned}
{\cal K}({\bf P},\z)=&\,\frac{\overline k(\z{\bf P})}{\z^n} + (-\z)^n k \left( -\frac{{\bf P}}{\z} \right )
\\
 =&\, \frac{\overline k\left ( {\bf \overline \Sigma} + \frac{{\bf \overline \Phi}}{\z}  
 + \z {\bf X} - \z^2 {\bf \Sigma} + \z^3 {\bf \Phi}  \right )}{\z^n}
  + (-\z)^n k \left( {\bf  \Sigma} - \frac{{\bf \overline \Phi}}{\z^3} - \frac{\overline{\bf \Sigma}}{\z^2} 
   -  \frac{{\bf X}}{\z} - \z {\bf \Phi}  \right )
 \, . 
\end{aligned}
\ee

\item[iii.] An expansion of ${\cal K}$ around ${\bf \Phi},\overline {\bf \Phi}$, 
namely 
\be
\begin{aligned}
\label{expPhi}
{\cal K}({\bf P},\z)=&\,\frac{\overline k(\z^2{\bf P})}{\z^n} + (-\z)^n k \left( \frac{{\bf P}}{\z^2} \right ) 
\\
=&\, \frac{\overline k\left ( {\bf \overline \Phi} + \z {\bf \overline \Sigma} + \z^2 {\bf X} - \zeta^3 {\bf \Sigma} 
+ \zeta^4 {\bf \Phi}  \right )}{\z^n} + (-\z)^n k \left( {\bf \Phi} + \frac{{\bf \overline \Phi}}{\z^4} 
+ \frac{{\bf \overline \Sigma}}{\z^3} + \frac{{\bf X}}{\z^2} - \frac{{\bf  \Sigma}}{\z}    \right ) 
\, . 
\end{aligned}
\ee

\end{enumerate} 
It turns out that the options i. and ii. do not give an interacting theory once plugged in \eqref{ansatz}. Therefore, the 
only option iii. is interesting for our discussion.
It is worth mentioning that the action \eqref{FULL} 
can be written in terms of the projective superspace action
\be
S= \frac{1}{ 32 \pi \ri}\int d^4 x \; \oint_C {d \z } \;  
 \z  \tD^2 \btD^2 
\left [ \frac{\overline k(\z^2{\bf P})}{\z^4} +  \z^4 k \left( \frac{{\bf P}}{\z^2} \right )
+\frac{ \overline v(\z^2{\bf P})}{\z^2} + \z^2 v \left( \frac{{\bf P}}{\z^2} \right ) \right ] \, ,
\label{O4-proj-model}
\ee
where $f(\Phi) = k'(\Phi)$, $W(\Phi) = v'(\Phi)$ and $\a=\b=\g=\mu=0$.
The case parametrized by the first two terms was introduced in the undeformed case in \cite{GHK}
while the last two terms are new.
From now on we will restrict to models described by \eqref{O4-proj-model}.

\subsection{The goldstino in the complex linear superfield}  

\label{GinCL}

In this subsection we work with the deformation of the form 
\be
\beta \ne 0 \, , \quad \alpha = \gamma = \mu = 0 \, . 
\ee
Within this setup the superfields $\Phi$, $\Sigma$ and $X$ are standard  ${\cal N}=1$ superfields 
defined by the constraints \eqref{thegood}, \eqref{thebad} and \eqref{theugly} respectively, 
and the $\rho$-supersymmetry is 
\be
\label{beta} 
\delta_{\rho} \Phi = - \overline \rho^{\dot \alpha} \Dd_{\dot \alpha} \Sigma  \, , \quad 
\delta_{\rho} \Sigma  = \rho^\alpha \D_\a \Phi
- \overline  \rho^{\dot \alpha} \Dd_{\dot \alpha} X  
+ \b \, \overline  \rho^{\dot \alpha} \overline \theta_{\dot \alpha} \, , \quad 
\delta_{\rho} X  = \rho^\alpha \D_\a \Sigma + \overline \rho^{\dot \alpha} \Dd_{\dot \alpha} \overline \Sigma 
\, . 
\ee 
The $\epsilon$-supersymmetry is preserved while the $\rho$-supersymmetry is broken and 
the goldstino belongs to the complex linear superfield. 
The lowest scalar component field of the complex linear multiplet has to possess a shift symmetry, 
therefore we require that the Lagrangians we write down have a shift symmetry on the superfield level, 
namely 
\be
\label{Sshift}
\Sigma \to \Sigma + \text{const.} 
\ee

 Let us now construct Lagrangians with two derivatives.
First we consider the action \eqref{FULL} of the undeformed theory, 
which is clearly not invariant under \eqref{beta}, 
and we impose the shift symmetry \eqref{Sshift}, 
which we know has to be respected by the deformed theory. 
This requirement implies 
\be
\label{betaf}
f(\Phi) = \frac12 \Phi  \, . 
\ee
The second step is to perform a supersymmetry transformation \eqref{beta} on \eqref{FULL} (with $f'=1/2$), 
which gives
\be
\label{bL}
\delta_{\rho} S = \beta \, [ \cdots ] + {\rm c.c.} 
\ee
Then we have to find a suitable compensating term which is both 
${\cal N}=1$ supersymmetric and also cancels the $\beta$ terms in \eqref{bL}. 
Indeed, 
it turns out that a compensating term does exist and it comes in the form of a superpotential: 
$\b \, W(\Phi)$. 
Once we put everything together we have the $\N =1$ action of the deformed theory 
\be
\label{hyp}
\begin{aligned}
S = & \int d^4x d^4 \theta  \Big{[} \Phi \overline \Phi 
+ \frac12 X^2 - \Sigma \overline \Sigma \Big{]} 
+ \Big{[}  \b \int d^4x d^2 \theta \, W(\Phi) + {\rm c.c.}   \Big{]} 
\\
& + \int d^4x d^4 \theta \Big{[} 
\frac12 W'(\Phi) \Sigma^2 + \frac12 \overline W'(\overline \Phi) \overline \Sigma^2 
+ X \left( W(\Phi) + \overline W(\overline \Phi) \right)
\Big{]} \, . 
\end{aligned}
\ee
The action \eqref{hyp} describes partial supersymmetry breaking with a complete ${\cal N}=2$ multiplet. 
Note that the above action is well-defined only for $W(\Phi) \ne \Phi$.

An interesting application which we can consider right away is to derive the model of \cite{GonzalezRey:1998kh} 
which contains only an ${\cal N}=1$ complex linear multiplet. 
In \cite{GonzalezRey:1998kh} the broken supersymmetry is non-linearly realized
and described by a nilpotent Goldstone multiplet, 
therefore to reduce to that model our ${\cal N}=2$ multiplet has to be appropriately truncated to an ${\cal N}=1$ 
complex linear superfield. 
This can be done by introducing a large mass for the chiral multiplet $\Phi$ and then 
decoupling it from our action \eqref{hyp}. 
To achieve this we set 
\be
W = m \Phi^2 \, , 
\ee
for a real constant $m$, 
which then brings \eqref{hyp} to the form 
\be
\label{hypm}
\begin{aligned}
S = & \int d^4x d^4 \theta  \Big{[} \Phi \overline \Phi 
+ \frac12 X^2 - \Sigma \overline \Sigma \Big{]} 
+ \Big{[} \b \int d^4x d^2 \theta \, m \Phi^2  + {\rm c.c.}   \Big{]}  
\\ 
& + \int d^4x d^4 \theta \Big{[} 
m \Phi  \Sigma^2 + m \overline \Phi  \overline \Sigma^2 + X \left( m \Phi^2  + m \overline \Phi^2  \right) 
\Big{]} \, . 
\end{aligned}
\ee
We see that $m$ is related to the mass of the chiral superfield. 
From \eqref{hypm} we derive the superspace equations of motion of $\Phi$ which read 
\be
\Dd^2 \overline \Phi + 2  \b m \Phi + m \Dd^2 \Sigma^2 + 2 m \Phi \Dd^2 X= 0 \, , 
\ee
and, assuming that  $ \langle \overline \D^2 X\rangle\ne - \b $,
we can recast them in the form 
\be
\label{EQphi}
\Phi  = \frac{- \overline \D^{\dot \alpha} \Sigma \, \overline \D_{\dot \alpha} \Sigma
 - m^{-1} \overline \D^2 \overline \Phi }{2 \, (  \b + \overline \D^2 X)} \, . 
\ee
We then consider the formal limit 
\be
\label{minf}
m \to \infty \, , 
\ee
which essentially gives an infinite mass to the full chiral multiplet $\Phi$ and therefore decouples it. 
With this procedure the superspace equation of motion \eqref{EQphi} turns into a constraint and a non-linear 
realization described by a nilpotent chiral  superfield can emerge. 
Indeed, 
in the formal limit \eqref{minf}, 
the superfield $\Phi$ will decouple and its equations of motion \eqref{EQphi} become 
\be
\label{RGR}
\Phi  = -\frac{1}{2} \frac{\Dd^{\dot \alpha} \Sigma \, \Dd_{\ad} \Sigma}{\b + \Dd^2 X} \, . 
\ee
Equation \eqref{RGR} is exactly the constraint arising in \cite{GonzalezRey:1998kh} 
for the study of partial supersymmetry breaking with 
the $\rho$-supersymmetry non-linearly realized and it implies $\Phi^2=0$.

To further study the properties of \eqref{hyp} we can integrate out $X$ directly from superspace which gives 
\be
X = - W(\Phi) - \overline W(\overline \Phi) \, , 
\ee 
and results in an action in terms of the chiral and the complex linear superfields only
\be
\label{hyp2}
\begin{aligned}
S = & \int d^4x d^4 \theta  \Big{[} \Phi \overline \Phi 
- |W(\Phi)|^2 - \Sigma \overline \Sigma 
+ \frac12 W'(\Phi) \Sigma^2 + \frac12 \overline W'(\overline \Phi) \overline \Sigma^2  \Big{]} 
\\&
 + \Big{[} \b \int d^4x d^2 \theta \, W(\Phi) + {\rm c.c.}   \Big{]}  \, . 
\end{aligned}
\ee 
The $\rho$-supersymmetry of \eqref{hyp2} is 
\be
\label{beta2} 
\delta_{\rho} \Phi = - \overline \rho^{\dot \alpha} \Dd_{\dot \alpha} \Sigma  \, , \quad 
\delta_{\rho} \Sigma  = \rho^\alpha \D_\a \Phi
+ W'(\overline \Phi) \, \overline  \rho^{\dot \alpha} \Dd_{\dot \alpha} \overline \Phi 
+  \b \, \overline  \rho^{\dot \alpha} \overline \theta_{\dot \alpha} \, . 
\ee 
By defining the components of the chiral superfield as in \eqref{componentsPhi} 
and for the complex linear as 
\be
\begin{aligned}
& \Sigma| = B \, ,  
\quad \D^2 \S | = C \, , 
\quad \overline \D_{\dot \alpha} \D_\alpha \S | = P_{\alpha \dot \alpha}   
\, ,
\\ 
& \overline \D_{\dot \alpha} \S | = \overline \tau_{\dot \alpha} \, , 
\quad \D_\alpha \S | = \nu_\alpha \, , 
\quad \frac12 \D^\gamma \overline \D_{\dot \alpha} \D_\gamma \S | = \overline \sigma_{\dot \alpha} 
\, ,
\end{aligned}
\ee
one can write down the action \eqref{hyp2} in component form and verify
 that indeed there are two complex scalars $A$ and $B$
and two Weyl fermions propagating. 
It is easier however to study the theory in the dual form in terms of two chiral superfields.
The details of this analysis are given  in appendix B. 
The vacuum structure of the theory and the preserved supersymmetry 
can be easily studied directly from \eqref{hyp2} 
by simply calculating the scalar potential which reads 
\be
{\cal V} = \beta^2  \,  \frac{W'(A) \overline W'(\overline A)}{1 - W'(A) \overline W'(\overline A)} \, . 
\ee
From the form of the scalar potential we can see that the vacua of the theory are always given by 
\be
\label{VACUUM}
\langle W' \rangle = 0 \, .
\ee 
The supersymmetry transformations of the fermions are given by 
\be
\begin{aligned}
\label{DF}
\delta \chi_\alpha & = \beta \, \frac{\overline W'}{1 - |W'|^2} \, \epsilon_\alpha + \text{terms with derivatives} \, , 
\\
\delta \overline \tau_{\dot \alpha} & =
 - \beta \, \overline \rho_{\dot \alpha} \left( 1 - \frac{|W'|^2}{1 - |W'|^2} \right) + \text{terms with derivatives}  \, , 
\\
\delta \nu_\alpha & =  \beta \, \frac{\overline W'}{1 - |W'|^2} \, \rho_\alpha + \text{terms with derivatives} \, , 
\\
\delta \overline \sigma_{\dot \alpha} & =  \text{only terms with derivatives} \, .
\end{aligned}
\ee  
From the transformations of the fermions \eqref{DF} we conclude that the theory always breaks supersymmetry 
partially, 
because of the vacuum condition \eqref{VACUUM}. 
Notice that if $W'$ was a constant both supersymmetries would be spontaneously broken. 
On the other hand, thanks to \eqref{VACUUM}, the pattern of partial breaking is the same 
as the one induced by the $\b$-deformation of the $\cN=2$ supersymmetry.

In parallel to the previous $\cN=1$ discussion, our projective superspace method  applies as follows.
Due to the central charge acting nontrivially on $\S$ the obvious candidate for ${\cal K}$ is \eqref{expPhi}. 
Repeating the same arguments as in \eqref{chirality}, \eqref{power}, we find that the functions
\be
\begin{aligned}
{\cal K}({\bf P},\z)&= \frac{\overline v(\z^2{\bf P})}{\z^2} + \z^2 v \left( \frac{{\bf P}}{\z^2} \right ) 
\, , 
\\
{\cal R}\left ( {\bf P},\z \right )&= 2\z^{-1}  \pa_{z} \left [   \z^2 v \left( \frac{{\bf P}}{\z^2}  \right ) 
-\frac{1}{\z^2}  \overline{v} \left( \z^2{\bf P}  \right ) \right ] 
\, , 
\end{aligned}
\ee 
give an $\N =2$ invariant action 
when inserted into \eqref{newaction}. However, there is no part in ${\cal K}({\bf P},\z)$ 
that corresponds to the free kinetic action. 
Therefore we add a quadratic term to ${\cal K}({\bf P},\z)$. Due to \eqref{geta2} 
we need to compensate its transformation by adding an appropriate ${\cal R}_{{\bf P}^2}$. Finally, we have
\be
\begin{aligned}
\label{antonfgO4}
{\cal K}({\bf P},\z)&= \frac{{\bf P}^2}{2} + \frac{\overline v(\z^2{\bf P})}{\z^2} 
+ \z^2 v \left( \frac{{\bf P}}{\z^2} \right ) 
\, , 
\\
{\cal R}\left ( {\bf P},\z \right )&= 2\z^{-1}  \pa_{z} \left [ \frac{{\bf P}^2}{2} 
+  \z^2 v \left( \frac{{\bf P}}{\z^2}  \right ) 
-\frac{1}{\z^2}  \overline{v} \left( \z^2{\bf P}  \right ) \right ] 
\, , 
\end{aligned}
\ee
 which, once we use \eqref{SAA-b},
 leads exactly to \eqref{hyp} (for $v'(\Phi)=W(\Phi)$), 
which we derived earlier.
Therefore the two methods for finding the two-derivative Lagrangians are consistent.

Let us finally give some examples for higher-derivative interactions. 
As we have explained before, there is a large variety of such interactions one can write down. 
Moreover, as we have already explained, 
here we just construct these terms as a means to exemplify our method. 
For a better understanding of their properties a detailed study of their vacuum structure is required 
which is however beyond the scope of this work.

As a first example we can consider an interaction term of form 
\be
\begin{aligned}
\label{Xtheory}
S_{{\rm int.}1}= {1 \over 32 \pi \ri}\frac{1}{\Lambda^4} \int d^4 x  \oint_C {d \z } \; 
 \z  \tD^2 \btD^2  \, \Big [~& \frac{1}{\z^2} \tna^2 \btna^2 \left ( {\bf X}^4   \right ) \Big ]
 \Big \vert    \, , 
\end{aligned}
\ee 
which, once we use \eqref{SAA-b}, gives the corresponding $\cN=1$ action
\be
\begin{aligned}
S_{{\rm int.}1} = \frac{1}{2\Lambda^4} \int d^4x d^4 \theta \; 
   \Big {[}~  &6 \D^\a \S \D_\a \S \Dd^\ad \Sd \Dd_\ad \Sd + 12 X \D^2 \Phi \Dd^\ad \Sd \Dd_\ad \Sd  
\\
& 
+ 24 X \D^\a \S \Dd^\ad \D_\a X \Dd_\ad \Sd 
- 12 X^2 \Dd^\ad \D^2 \S \Dd_\ad \Sd
\\
& 
+ 6 X^2 \Dd^\ad \D^\a X \Dd_\ad \D_\a X
+ 12 X \D^\a \S \D_\a \S \Dd^2 \overline \Phi
\\
& 
+ 12 X^2 \D^2 \Phi \Dd^2 \overline \Phi
- 12 X^2 \D^\a \S  \Dd^2 \D_\a \Sd 
+ 4 X^3 \Dd^2 \D^2 X
    \Big{]}
    +{\rm c.c.}
\end{aligned}         
\ee
Another example is 
\be
\begin{aligned}
S_{{\rm int.} 2}=  {1 \over 32 \pi \ri}\frac{1}{\Lambda^4} \int d^4 x \oint_C {d \z } \; 
 \z  \tD^2 \btD^2  \, \Big [~&  \frac{1}{\z^2} \tna^2 \btna^2 \left ( {\bf \Phi}^2 \overline {\bf \Phi}^2  \right ) \Big ]
 \Big \vert    \, , 
\end{aligned}
\ee 
which in $\N=1$ superspace takes the form 
\be
\begin{aligned}
S_{{\rm int.} 2} = \frac{1}{2\Lambda^4} \int d^4x d^4 \theta \; 
   \Big [~  & \pa^{\a \ad} \Phi  \pa_{\a \ad} \Phi \overline \Phi^2 
   + 2 \Phi \Box \Phi \overline \Phi^2
+ 4 \ri  \pa^{\a \ad} \Phi \Dd_\ad \S \D_\a \Sd  \overline \Phi  
+ 2 \b \Dd^\ad \S \Dd^\ad  \S \overline \Phi
\\
&+ 2  \b \D^\a \Sd \D^\a  \Sd  \Phi  
+ 2  \Dd^\ad \S \Dd_\ad \S \D^2 X \overline \Phi
+ 2  \D^\a \Sd \D_\a \Sd \Dd^2 X  \Phi
\\
&+ 4 \b^2 \Phi \overline \Phi
+\Dd^\ad \S \Dd_\ad \S \D^\a \Sd \D_\a \Sd
+ 4 \ri \Phi \pa^{\a \ad} \Dd_\ad \S \D_\a \Sd \overline \Phi 
\\
&+ 4 \b \Phi \overline \Phi \left ( \Dd^2 X
+ \D^2 X \right )
+ 4 \Phi \overline \Phi \Dd^2 X \D^2 X
    \Big]
         +{\rm c.c.}
\end{aligned}         
\ee

\subsection{The goldstino in the chiral superfield}

\label{GinC}

In this subsection we study the deformation 
\be
\mu \ne 0 \, , \quad \alpha = \beta = \gamma = 0 \, . 
\ee
In this setup  
the $\rho$-supersymmetry is 
\be
\label{mu} 
\delta_{\rho} \Phi = - \overline \rho^{\dot \alpha} \Dd_{\dot \alpha} \Sigma  
- \mu \, \rho^\alpha \theta_\alpha  
\, , \quad 
\delta_{\rho} \Sigma  = \rho^\alpha \D_\a \Phi
- \overline  \rho^{\dot \alpha} \Dd_{\dot \alpha} X  
\, , \quad 
\delta_{\rho} X  = \rho^\alpha \D_\a \Sigma + \overline \rho^{\dot \alpha} \Dd_{\dot \alpha} \overline \Sigma 
\, . 
\ee 
The $\epsilon$-supersymmetry is preserved while the $\rho$-supersymmetry is broken and the goldstino belongs to 
the ${\cal N}=1$ chiral superfield $\Phi$. 
The lowest scalar component field of the chiral multiplet has to possess a shift symmetry, 
therefore we require that the Lagrangians we write down have a shift symmetry at the superfield level, 
namely 
\be
\label{Fshift}
\Phi \to \Phi + \text{const.} 
\ee
This restricts the holomorphic functions in \eqref{FULL} to the form 
\be
f(\Phi) = \frac12 \Phi \, , \quad W = c \, , 
\ee
where $c$ is a complex constant. 
The two-derivative theory then takes the form 
\be
\label{Lmu}
S = \int d^4x d^4 \theta \left( \Phi \overline \Phi +\frac12 X^2 - \Sigma \overline \Sigma \right) \, . 
\ee 
The model breaks supersymmetry partially and the fields form a complete ${\cal N}=2$ multiplet.  
This is easily seen by reducing the theory to components and checking the transformations of the fermions.

The projective superspace method used to find lagrangians with no more
 than two derivatives gives \eqref{Lmu} in agreement with the more intuitive method used to derive it.
Therefore, in order to find nontrivial interaction terms 
we need to use lagrangians with higher derivatives as in \eqref{SAA}. 
As an example we give
\be
\begin{aligned}
S_{\rm int}= \frac{1}{\Lambda^4} {1 \over 32 \pi \ri} \int d^4 x \oint_C {d \z } \; 
 \z  \tD^2 \btD^2  \, \Big [~&\frac{1}{\z^2} \tna^2 \btna^2 \left ( {\bf \S}^2 \overline {\bf \S}^2  \right ) \Big ]
 \Big \vert    \, , 
\end{aligned}
\ee 
which in ${\cal N}=1$ superspace, once we use \eqref{SAA-b},  reads
\begin{eqnarray} 
\nn
\break 
S_{\rm int.} =\frac{1}{2\Lambda^4} \int d^4x d^4 \theta \; 
   \Big[ 
   && \hspace{-0.6cm} 
   2 \ri \pa^{\a \ad} \D_\a  \Phi \Dd_\ad X \Sd^2
+ \Dd^\ad \D_\a \S \Dd_\ad \D^\a \S \Sd^2 
+ 4  \Dd^\ad \D^\a \S \Dd_\ad X \D_\a X \Sd 
\\
\nn
+ &&\hspace{-0.6cm}  2 \Sd \Dd_\ad X \Dd^\ad  X \D^2 \S 
+  \Dd^\ad X \Dd^\ad  X  \D^\a X  \D_\a X
+ \Dd^\ad \overline \Phi \Dd_\ad \overline \Phi \D^\a \Phi \D_\a \Phi
\\
\nn
+ &&\hspace{-0.6cm} 2  \mu \S  \Dd^\ad \overline \Phi \Dd_\ad \overline \Phi
+4 \S \D^\a \Phi \Dd_\ad \overline \Phi  \Dd^\ad \D_\a \Sd 
+ 2 \S^2 \Dd^\ad \D^2 X \Dd_\ad \overline \Phi
\\
\nn
+ &&\hspace{-0.6cm}  \S^2  \Dd^\ad \D_\a \Sd   \Dd_\ad \D^\a \Sd
- 2  \mu \S^2 \D^2 \S 
+2 \Sd^2 \D^\a \Phi \Dd^2 \D_\a X 
\\
+ &&\hspace{-0.6cm} 4 \S \Sd \Dd^2 \D_\a X \D^\a X
+4  \Sd \D^\a \Phi \D_\a X  \Dd^2 \Sd 
+ 4 \S \Sd \Dd^2 \Sd \D^2 \S
\\
\nn
+&&\hspace{-0.6cm} 2 \S \D_\a X \D^\a X \Dd^2 \Sd 
+ 4 \mu^2 \S \Sd 
+  2 \mu \Sd \D^\a \Phi \D_\a \Phi 
\\
\nn
+&&\hspace{-0.6cm} 4  \mu \S \D_\a \Phi  \D^\a X 
+  4 \mu \Sd \Dd^\ad X \Dd_\ad X
+ 2 \S \Sd^2 \Dd^2 \D^2 \S
+ 2 \mu^2 \Sd^2
\\
\nn
+ &&\hspace{-0.6cm} 4 \D^\a \Phi \D_\a X \Dd^\ad X \Dd_\ad \overline \Phi
+ 4 \Sd \D^\a \Phi  \Dd_\ad \overline \Phi \Dd^\ad \D_\a \S 
+  4 \S \D^2 \S \Dd^\ad X \Dd_\ad \overline \Phi
\\
\nn
+ &&\hspace{-0.6cm}  4 \S \D^\a X \Dd^\ad \overline \Phi \Dd_\ad \D_\a \S 
+ 4\ri \S \Sd  \pa_\a^{~\ad} \D^\a \Phi \Dd_\ad \overline \Phi
+ 4 \Sd \D^\a \Phi  \Dd^\ad X \Dd_\ad \D_\a \Sd
\\
\nn
+&&\hspace{-0.6cm} 4 \S \D^\a X \Dd_\ad X \Dd^\ad \D_\a \Sd
+4 \S \Sd \Dd^\ad \D_\a \S \Dd_\ad \D^\a \Sd
+4 \S \Sd \Dd_\ad X \Dd^\ad \D^2 X
    \Big]+{\rm c.c.}
\end{eqnarray}         
Another possible higher-derivative interaction for the current setup is given by \eqref{Xtheory}. 
Again we have not investigated the vacuum structure or the possible presence of ghost-like excitations
introduced by this term.

\section{Discussion} 
\label{Discussion}

In this paper we have introduced new models giving rise to global $4D$ partial supersymmetry breaking 
from ${\cal N}=2$ to ${\cal N}=1$.
By considering self-interacting off-shell hypermultiplets,  
the main idea of our paper was to characterize the patterns of supersymmetry breaking in terms 
of the broken central charge symmetries that lead to deformed $\cN=2$ supersymmetry algebras.
To develop this idea, we used ${\cal N}=2$ projective superspace with central charges and 
systematically studied constant deformations of $\cO(2)$ and $\cO(4)$ multiplets.
Projective superspace is known to be eminently suited for the purpose of reducing manifestly 
off-shell $\cN=2$ theories to
an $\cN=1$ superspace description. As such, it has proven to be a natural setup to describe models for 
${\cal N}=2$ to ${\cal N}=1$ supersymmetry breaking.
 Within our approach we reproduced the previously known results for partial supersymmetry breaking
 of \cite{Rocek:1997hi,GonzalezRey:1998kh,Antoniadis:2017jsk,Kuzenko:2017gsc}
 and we described new models with and  without higher-derivative interactions.

The analysis in our paper opens the venue for various generalizations and new research that we plan to pursue
in the future.
In particular, the nontrivial examples of higher-derivative models we have constructed in this paper
 deserve a more thorough analysis. 
For instance, 
we have left for 
the future a detailed study of the vacua, which might modify the supersymmetry breaking patterns,
and the existence of ghost modes in these models. 
It is also of interest to describe new higher-derivative actions possessing deformed $\cN=2$ supersymmetry
and in particular to look for possible ghost-free models generalizing the known $\cN=1$ results of
 \cite{Khoury:2010gb,Koehn:2012ar,Farakos:2013zsa,Farakos:2014iwa,Farakos:2015vba}.

In this paper we have focused on real $\cO(2)$ and $\cO(4)$ hypermultiplets.
A natural generalization is to extend our analysis to self interacting \emph{polar} multiplets \cite{G-RRWLvU}.
Similarly to the $q^+$ hypermultiplet in harmonic superspace \cite{GIKOS, GIOS},
the polar multiplet provides a fully off-shell formulation of the charged hypermultiplet.
The polar multiplet is described in terms of the so-called arctic superfield, 
$\U=\sum_{k=0}^{+\infty}\z^k \U_k$,
which can be seen as a $k\to \infty$ limit of a complex $\cO(k)$ multiplet and as such
it contains an infinite number of $\cN=1$ superfield components. By extending the setup of our paper it might be 
possible to describe deformed arctic multiplets with an infinite set of $(\pa_z \U_k=\a_k,\pa_{\bar{z}} \U_k=\b_k)$ 
constant deformation 
parameters that might describe new nontrivial patters of $\cN=2\to\cN=1$ supersymmetry breaking.
It might also be interesting to explore the possibility to start from  non-constant deformed supersymmetry and 
central charge transformations. In this case the $(\a_k,\b_k)$ 
might be nontrivial functions of $\cN=1$ superfields 
and be eventually related to broken isometries in the target space geometry
of $\cN=2$ sigma-models.

Another natural question concerns the extension of our analysis beyond hypermultiplets.
As a matter of fact the first model of $4D$ partial supersymmetry breaking by Antoniadis--Partouche--Taylor  (APT)
was based on magnetically deformed interacting $\cN=2$ vector multiplets \cite{Antoniadis:1995vb}.
Since an $\cN=2$ vector multiplet is decomposed in an $\cN=1$ chiral scalar and a vector multiplet
it is not difficult to realize that the approach pursued in our paper cannot be applied straightforwardly
to reproduce the original APT model.  
In fact, the deformed algebra in the APT case includes a vector charge (not only scalar charges) that extends 
the standard $\cN=2$ Poincar\'e superalgebra.  
In a phase that preserves $\cN=1$ supersymmetry,
in the deformed $\cN=2$ vector multiplet of the APT model, the $\cN=1$ vector multiplet 
is the one possessing the Goldstone modes, see e.g. \cite{IZ} for a nice discussion on this subject. 
It would be interesting to modify our setup to allow for more general central extensions of the $\cN=2$ Poincar\'e
algebra.
In particular, this naturally leads to the study of $5D$ and $6D$ models in terms of $4D$ superfields
as for instance in \cite{Kuzenko:2005sz,GPT-M}.

It would also be  interesting to investigate whether the ideas in our paper can be extended to the case of 
partially broken local supersymmetry
and relate our new models to the global limit of a specific ${\cal N}=2$ supergravity theory.  
For example, 
it is natural to argue 
that  
the new model introduced in section 2 (and the model of subsection \ref{GinC}) 
might arise from  
a hypermultiplet sector of one of the supergravity models already discovered in the past. 
The supergravity theory we will be interested in is the one studied in \cite{Ferrara:1995xi} 
and more recently in \cite{Antoniadis:2018blk} 
where it is shown how the APT model \cite{Antoniadis:1995vb} can be recovered in the global limit. 
We argue that the partial breaking related to the deformations \eqref{ChiralDef} 
is in fact dual to the hidden hypermultiplet sector of \cite{Ferrara:1995xi}. 
Indeed in the global limit in \cite{Ferrara:1995xi} the hypermultiplet sector decouples from the $\cN=2$
 vector multiplet 
and from supergravity. 
The bosonic matter sector of this theory contains the four real scalars of the hypermultiplet $q^u$ (with $u=0,1,2,3$), 
a complex scalar $z$ belonging to the vector multiplet and the abelian gauge field of the vector multiplet. 
The scalar potential will depend on $z$ and $q^0$, 
however for specific values of the couplings (see \cite{Ferrara:1995xi} and \cite{Antoniadis:2018blk}), 
the complex scalar $z$ gets stabilized 
irrespective of the value of $q^0$ 
such that 
\be
\label{VG}
{\cal V} ~ \Big{\vert}_{z=\langle z \rangle} \equiv 0 \, ,  
\ee 
yielding  ${\cal N}=1$ vacua. 
The full supergravity scalar potential ${\cal V}$ can be found in \cite{Antoniadis:2018blk} 
where its properties are studied in detail. 
Moreover, 
in the global limit the hypermultiplet scalars have kinetic terms 
\be
\label{KG}
- \frac{M_P^2}{(M_P + q^0)^2} \delta_{uv} \,\partial_m q^u\partial^m q^u
= - \delta_{uv}\,\partial_m q^u\partial^m q^u  + {\cal O}\left(q^0 /M_P \right)   \, , 
\ee 
therefore the hyper-K\"ahler metric is flat. 
The form of the scalar potential \eqref{VG}, 
the form of the kinetic terms \eqref{KG}, 
and the fact that supersymmetry is partially broken by the hypermultiplet sector imply that the new models 
in section \ref{REAL} and section \ref{O4} 
might be related to the global limit of the hidden sector of \cite{Ferrara:1995xi}. 
Extensions of this setup can be found in \cite{Andrianopoli:2015rpa},
see also the recent analysis of \cite{Antoniadis:2018blk}, 
but it is worth to mention that in these setups the hypers are all in an on-shell formulation. 
To have a more conclusive answer about our previous statements, 
it would be necessary to obtain a fully off-shell extension of the results of
\cite{Ferrara:1995xi} and \cite{Antoniadis:2018blk}, 
and then analyze in detail which off-shell hypermultiplets can support these models. 
This might be possible by using the covariant projective superspace approach of 
\cite{ProjectiveSugra4D}. See \cite{Kuzenko:2015rfx,Kuzenko:2017zla,Kuzenko:2017gsc} for recent application of 
this approach to broken $\cN=2$ supersymmetry both in flat and curved backgrounds. 
If the extension is only based on a system of off-shell vector and tensor multiplets,
then the results of \cite{deWit:2006gn}, see also the rheonomic superspace description of \cite{Cribiori:2018xdy},
will also be important.
Then, by taking a rigid limit of a fully off-shell extension of  \cite{Ferrara:1995xi}, 
we could observe directly 
which set of auxiliary fields complete the hypermultiplet sector.

\section*{Acknowledgments}  

We thank Klaus Bering, Ariunzul Davgadorj, 
Jean-Pierre Derendinger, 
Ond\v{r}ej Hul\'{i}k, Konstantinos Koutrolikos, 
Sergei M. Kuzenko, Michal Pazderka and Konstantinos Siampos for useful discussions. 
F.F., P.K. and R.v.U. are grateful to the organizers of the 38th Geometry and Physics Winter School in
 Srn\'{i} for creating the wonderful atmosphere where part of this work was carried out. 
P.K. would like to thank the University of Padova for the kind hospitality where this project was initiated. 
This work was supported in part by the Interuniversity Attraction Poles Programme initiated 
by the Belgian Science Policy (P7/37), 
by the KU Leuven C1 grant ZKD1118 C16/16/005,
and by the grant P201/12/G028 of the Grant agency of Czech Republic.
G.T.-M. is also supported 
by the Albert Einstein Center for Fundamental Physics, University of Bern,
and by the Australian Research Council (ARC) Future Fellowship FT180100353.

\appendix

\section{The ${\cal N}=1$ deformed real linear multiplet} 
\label{AppendixA}

As we have seen in the bulk of the article, 
the new model for partial breaking described in section \ref{REAL} 
naturally gives rise to a deformed real linear superfield originally proposed in \cite{Kuzenko:2017oni}, 
which is a real superfield $L= \overline L$, 
satisfying the constraint 
\be
\label{ML}
\D^2 L = f = \Dd^2 L \, ,
\ee 
with $f$ a constant parameter that, for simplicity, we choose to be real.
In this section we describe how to uncover the modified supersymmetry transformations for a real linear multiplet 
and discuss some of its properties.

We start from a model with a single $\cN=1$ chiral superfield $\phi$ 
($\overline  \D_{\ad} \phi =0$) 
describing spontaneous supersymmetry breaking, 
and then we dualize to the real linear multiplet. 
To perform the duality the chiral model has to posses an isometry 
\be
\phi \to \phi + \ri c\, , 
\ee
where $c$ is a real constant. 
The simplest model is 
\be
\label{LL2}
{\cal L} = \frac12 \int d^4 \theta \, \left( \phi + \overline \phi \right)^2 
- \left( f \int d^2 \theta \, \phi + {\rm c.c.} \right) . 
\ee 
Here $f$ is a constant which we set to be real and it is related to the supersymmetry breaking scale. 
Once we write the Lagrangian \eqref{LL2} in component form we see that 
supersymmetry is broken because $\langle \D^2 \phi| \rangle \ne 0$, 
and it gives rise to a goldstino identified with the fermion in $\phi$. 
The Lagrangian \eqref{LL2} can be written as 
\be
\label{Ldual}
{\cal L}_\text{dual} = - \frac12 \int d^4 \theta \, L^2 + \int d^4 \theta \, \left( \phi + \overline \phi \right) L 
-  \left( f \int d^2 \theta \, \phi + {\rm c.c.} \right) , 
\ee
where now $L$ is a real but otherwise unconstrained superfield ($L=\overline{L}$). 
By integrating out $L$ from \eqref{Ldual} we find the Lagrangian \eqref{LL2}.

Now we integrate out $\phi$ to uncover the dual model. 
Indeed the variation of $\phi$ gives \eqref{ML}, 
and the Lagrangian \eqref{Ldual} becomes 
\be
\label{LL3}
{\cal L} = - \frac12 \int d^4 \theta \, L^2 . 
\ee
Now $L$ has become a deformed real linear superfield, 
and has component fields \eqref{Lcomp}.  
The Lagrangian \eqref{LL3} in component form reads 
\be
{\cal L} = -  f^2 +\frac{1}{2} t^{\a \ad} t_{\a \ad} 
+ \frac{1}{8}  l \pa^{\a \ad} \pa_{\a \ad} l 
+ \ri  \lambda^{\a} \pa_{\a}^{~ \ad} \overline  \lambda_{\ad} \, . 
\ee
Here supersymmetry is spontaneously broken and the goldstino is given by $\lambda_\alpha$, 
which transforms under supersymmetry as 
\be
\label{dpsi}
\delta \lambda_\alpha = - \epsilon_{\a} f - \overline \epsilon^{\ad} \left ( t_{\a \ad} - \frac{\ri}{2} \pa_{\a \ad} l \right ) . 
\ee
We see that the deformed real linear superfield arises from dualizing to the chiral superfield $\phi$
model  which breaks supersymmetry.

The real linear superfield we just presented has some properties which we would like to explore further. 
One could have directly considered a supersymmetric Lagrangian of the form 
\be 
\label{L111}
{\cal L} = \left( \frac14 \int d^2 \theta \, \overline{\D}_{\dot \alpha} L \,  \overline{\D}^{\dot \alpha} L  + {\rm c.c.} \right) , 
\ee 
which in component form gives 
\be
\label{LC111}
{\cal L} =  \frac{1}{2} t^{\a \ad} t_{\a \ad} 
+ \frac{1}{8}  l \pa^{\a \ad} \pa_{\a \ad} l 
+ \ri  \lambda^{\a} \pa_{\a}^{~ \ad} \overline  \lambda_{\ad}
~. 
\ee
Notice that here the fermion $\lambda_\alpha$ is still the goldstino because supersymmetry acts on it as \eqref{dpsi}, 
however the vacuum energy in \eqref{LC111} is zero. 
Therefore global supersymmetry is spontaneously broken with vanishing vacuum energy. 
Moreover, this model explicitly violates the $R$-symmetry due to the defining constraint \eqref{ML}. 
This is in sharp contrast to the Volkov--Akulov model 
\cite{VA}
which has an $R$-symmetry in the leading order terms, 
albeit higher order terms can explicitly break the $R$-symmetry.

The effect of supersymmetry breaking can be made manifest once we mediate the breaking to the matter sector. 
For example one can have a chiral multiplet $Y$ and consider the term 
\be
{\cal L}_{\rm med} = - \frac{m^2}{f^4} 
\int d^4 \theta \left( Y \overline Y \,  D^\alpha L D_\alpha L \,  
\overline D_{\dot \alpha} L \overline D^{\dot \alpha} L \right) \, , 
\ee
which leads to 
\be
{\cal L}_{\rm med} \sim - m^2 Y \overline Y +\cdots 
\, , 
\ee
where the ellipses indicate  
terms with goldstini, 
and therefore this term generates a non-supersymmetric mass for the complex scalar $Y$.

\section{Supercurrents}
\label{AppendixB}

Here we derive the supercurrents of the Lagrangian \eqref{LC1} under the $\rho$- and 
$\epsilon$-supersymmetries. 
To identify the supercurrents we preform local supersymmetry variations on the Lagrangian \eqref{LC1}, 
and then up to total derivatives we obtain 
\be
\begin{aligned}
\delta_{\epsilon} {\cal L}|_{\text{total derivative}} =&\, \pa^{\a \ad} \epsilon^\b(x) \, J_{\a \b \ad \, \1} \, , 
\\
\delta_{\rho} {\cal L}|_{\text{total derivative}} =&\, \pa^{\a \ad} \rho^\b(x) \, J_{\a \b \ad \, \2} \, , 
\end{aligned}
\ee
which gives 
\be
\begin{aligned}
J_{\a \b \ad \, \1} =&- \ri f \, C_{\a \b} \overline \lambda_\ad 
+ \ri t_{\b \ad} \lambda_\a - \frac{1}{2} \lambda_\a \pa_{\b \ad}l - \chi_\a \pa_{\b\ad} \overline A \, , 
\\
 J_{\a \b \ad \, \2} =& - \ri t_{\b \ad} \chi_\a - \frac{1}{2} \chi_\a \pa_{\b \ad}l + \lambda_\a \pa_{\b\ad} A \, . 
\end{aligned}
\ee
Notice that on the mass shell the supercurrents are conserved,
$\partial^{\a \ad} J_{\a \b \ad \, \1}=0$ and $\partial^{\a \ad} J_{\a \b \ad \, \2}=0$. 
For the supersymmetry of the supercurrents we find 
\cite{Hughes:1986fa,Hughes:1986dn,Bagger:1997me,Porrati:1996xu} 
\be
\label{Q1}
\Big{\langle} \left \{ \overline \Q^{\1}_\bd ,\ J_{\a \b \ad \, \1} \right \} \Big{\rangle} =  
 \ri \, C_{\a \b} \overline C_{\ad \bd} \, f^2 
 \, , \quad 
\Big{\langle} \left \{ \overline \Q^{\2}_\bd ,\ J_{\a \b \ad \, \2} \right \} \Big{\rangle} = 0 \, . 
\ee

\section{Dualities from $\cO(4)$ to double-chiral and to $\cO(2)$ multiplets} 
\label{AppendixC}

In this appendix we focus on some further properties of the model we presented in section \ref{GinCL}, 
by dualizing the complex linear superfield. 
We will show that such model is actually equivalent to the
deformed $\cO(2)$ models studied in \cite{Antoniadis:2017jsk} and in sections \ref{REAL} and \ref{O2}.

We first dualize the complex linear $\S$ to a chiral superfield $Y$. 
Therefore we consider the Lagrangian 
\be
\label{hyp2APP}
\begin{aligned}
{\cal L} = & \int d^4 \theta  \Big{[} \Phi \overline \Phi 
- |W(\Phi)|^2 - \Sigma \overline \Sigma 
+ \frac12 W'(\Phi) \Sigma^2 + \frac12 \overline W'(\overline \Phi) \overline \Sigma^2  \Big{]} 
\\
& 
+ \int d^4 \theta \Big{[} \Sigma \, Y + \overline \Sigma \, \overline Y \Big{]} 
+ \Big{[}  \b \int d^2 \theta \, W(\Phi) + {\rm c.c.}   \Big{]}  \, , 
\end{aligned}
\ee
where $Y$ is a chiral superfield and now $\Sigma$ is unconstrained. 
When we vary $Y$ we get back \eqref{hyp2} with $\Sigma$ a complex linear. 
From \eqref{hyp2APP} we can also integrate out the unconstrained complex $\Sigma$ which gives 
\be
\label{SY}
\Sigma = \frac{\overline Y + \overline W'(\overline \Phi) Y}{1 - |W'(\Phi)|^2} \, , 
\ee
and, when inserted back into \eqref{hyp2APP}, we find 
\be
\label{hyper}
\begin{aligned}
{\cal L} = &  \int d^4 \theta  \left[\,
|\Phi|^2 - |W(\Phi)|^2 
+  \frac{  |Y|^2 + \frac12 \overline W'(\overline \Phi) Y^2 + \frac12 W'(\Phi) \overline Y^2 }{1 - |W'(\Phi)|^2} \right]
\\
& +  \Big{[}  \b \int d^2 \theta \, W(\Phi) + {\rm c.c.}   \Big{]} \, . 
\end{aligned}
\ee 
Now we perform the redefinition 
\be
Y = S \left( 1 - W'(\Phi) \right)  \, , 
\ee 
for $S$ a chiral superfield, which brings the K\"ahler potential to the form 
\be
\label{KSF} 
K = |\Phi|^2 - |W(\Phi)|^2 - \frac12 (S - \overline S)^2 \frac{|1 - W'(\Phi)|^2}{1 - |W'(\Phi)|^2}  \, . 
\ee
Note that the K\"ahler potential \eqref{KSF} has the shift symmetry $S \to S + c$, 
for a real constant $c$, 
which essentially originates from the spontaneously broken central charge symmetry.

We can further dualize the chiral superfield $S$ in the model with the K\"ahler potential \eqref{KSF} 
and superpotential $\b \int d^2 \theta \, W(\Phi)$ 
to an equivalent model with a real linear superfield $G$ and of course with the chiral $\Phi$ superfield. 
By performing this duality we get 
\be
\label{LPm}
\begin{aligned}
{\cal L} 
= & \int d^4 \theta \left[\,  |\Phi|^2 - |W(\Phi)|^2  - \frac12 G^2 
\left( \frac{1}{1-W'(\Phi)} +\frac{1}{1-\overline W'(\overline \Phi)} - 1 \right)  \right] 
\\
& +  \left[  \b \int d^2 \theta \, W(\Phi) + {\rm c.c.}   \right] \, . 
\end{aligned}
\ee
The $\cO(2)$ form of the action 
can be found after performing a final field redefinition 
\be
\label{redef}
\Psi = \Phi - W(\Phi) \, , 
\ee 
where $\Psi$ is a chiral superfield. 
In principle, we would have to invert \eqref{redef} to find $\Phi = {\cal F}(\Psi)$ and replace in \eqref{LPm}. 
However, 
we do not really have to find the inverse function ${\cal F}(\Psi)$, 
but  by just assuming it exists we have 
\be
\Phi = {\cal F}(\Psi) \, , \quad W(\Phi) =  {\cal F}(\Psi) - \Psi \, , 
\ee
and 
\be
{\cal F}'(\Psi) = \frac{\partial \Phi}{\partial \Psi} \equiv \left( \frac{\partial \Psi}{\partial \Phi} \right)^{-1} 
= \frac{1}{1 - W'(\Phi)} \, . 
\ee
Once we insert these equations into \eqref{LPm} it takes the standard $\cO(2)$ form 
\be
\label{LPmredef}
\begin{aligned}
{\cal L} = &\int d^4 \theta \left[  \Psi  \overline{\cal F}(\overline \Psi) +  \overline \Psi {\cal F}(\Psi) - \Psi \overline \Psi   
- \frac12 G^2 \left(  {\cal F}'(\Psi) + \overline{\cal F}'(\overline \Psi) - 1 \right)  \right] 
\\
& +  \left[  \b \int d^2 \theta \, \left( {\cal F}(\Psi) - \Psi \right) + {\rm c.c.}   \right] \, , 
\end{aligned}
\ee
which is of the type of models studied in \cite{Antoniadis:2017jsk} and in sections \ref{REAL} and \ref{O2}.


\end{document}